\documentclass[aps,floatfix,varwidth,border=10pt]{revtex4}
\usepackage{graphicx,amssymb,amsfonts,amsmath,color,bm,hyperref,subfig}
\usepackage{pdfpages}
\usepackage{pdftexcmds}
\usepackage{xcolor}
\hypersetup{
    colorlinks,
    linkcolor={blue},
    citecolor={blue},
    urlcolor={black}
}

\usepackage{makeidx}
\usepackage{graphicx}
\usepackage{chemfig}
\usepackage[capitalise,nameinlink,noabbrev]{cleveref}
\begin{document}
\title{Small RNA driven   feed-forward loop: critical role of sRNA in noise filtering}

\author{Swathi Tej${}^*$,  Kumar Gaurav${}^{**}$ and Sutapa Mukherji${}^{*}$}
\affiliation{${}^*$Department of Protein Chemistry and Technology, 
CSIR - Central Food Technological Research Institute, 
Mysore 570 020, India\\
${}^{**}$ Department of Biosciences and Bioengineering, Indian Institute of Technology, Guwahati 781 039, India}

\date{\today}
\begin{abstract}
Gene regulatory networks are often partitioned into different types of recurring network motifs. A feed-forward loop (FFL) is a common motif in which an 
upstream regulator is a protein, typically a transcription factor, that regulates the expression of the target protein in two ways - first, directly by regulating the 
mRNA levels of the target protein and second, indirectly via an intermediate molecule that in turn regulates the target protein level. Investigations on two variants of FFL - 
purely transcriptional FFL (tFFL) and sRNA-mediated FFL (smFFL) reveal several advantages of using such motifs. Here, we study a distinct sRNA-driven FFL (sFFL) 
that was discovered recently in {\it Salmonella enterica}: The distinction being the upstream regulator here is not a protein but an sRNA that translationally 
activates the target protein expression directly; and also indirectly via regulation of the transcriptional activator of the target protein. This variant, i.e. sFFL has not been subjected to rigorous analysis. We, therefore, set out to understand two aspects. First is a quantitative comparison of the regulatory response of sFFL with tFFL and smFFL using a differential equation framework. Since the process of gene expression is inherently stochastic, the second objective is to find how noise in gene expression affects the functionality of the sFFL. We find that unlike for tFFL and smFFL, the response of sFFL is stronger and faster: the change in target protein concentration is rapid and depends critically on the initial concentration of sRNA. Further, our analysis based on  generating function approach and stochastic simulations leads to a non-trivial prediction that an optimal noise filtration can be  attained depending on the synthesis rate of the upstream sRNA and the degradation rate of the intermediate transcriptional activator. A comparison with a simpler process involving only  translational activation by sRNA indicates that  the design of sFFL is  crucial 
for optimal noise filtration. These observations prompt us to conclude that sFFL has distinct advantages where the master regulator, sRNA, plays a critical role not only 
in driving a rapid and strong response, but also a reliable response that depends critically on its concentration.
\end{abstract}

\maketitle


\section{Introduction}
Small non-coding RNAs (small RNAs/micro RNAs) and proteins are two important regulators of  gene expression at different levels. While a 
major role of protein regulators is to activate or repress gene expression at the transcriptional level by binding to DNA, the small RNAs (sRNA) often 
bind to their target mRNAs through sequence complementarity and regulate translation. Experimental studies indicate that sRNAs might regulate gene 
expression in different ways - by translational repression or activation, or by regulating the mRNA stability {\cite{storz,gottesman2, papenfort1,altuvia,shimoni,hwa,bharat}}. 
Since sRNAs are small in size and do not further code for a protein, it is believed that sRNA-mediated regulation can lead to quick response with less 
energetic cost compared to protein-mediated regulation.  
Several recent studies show that  the  gene expression often involves dual strategies combining both protein- and sRNA-mediated regulation.{\cite{shimoni,bose2}}

\begin{figure*}[h!]
 	\includegraphics[width=0.35 \textwidth]{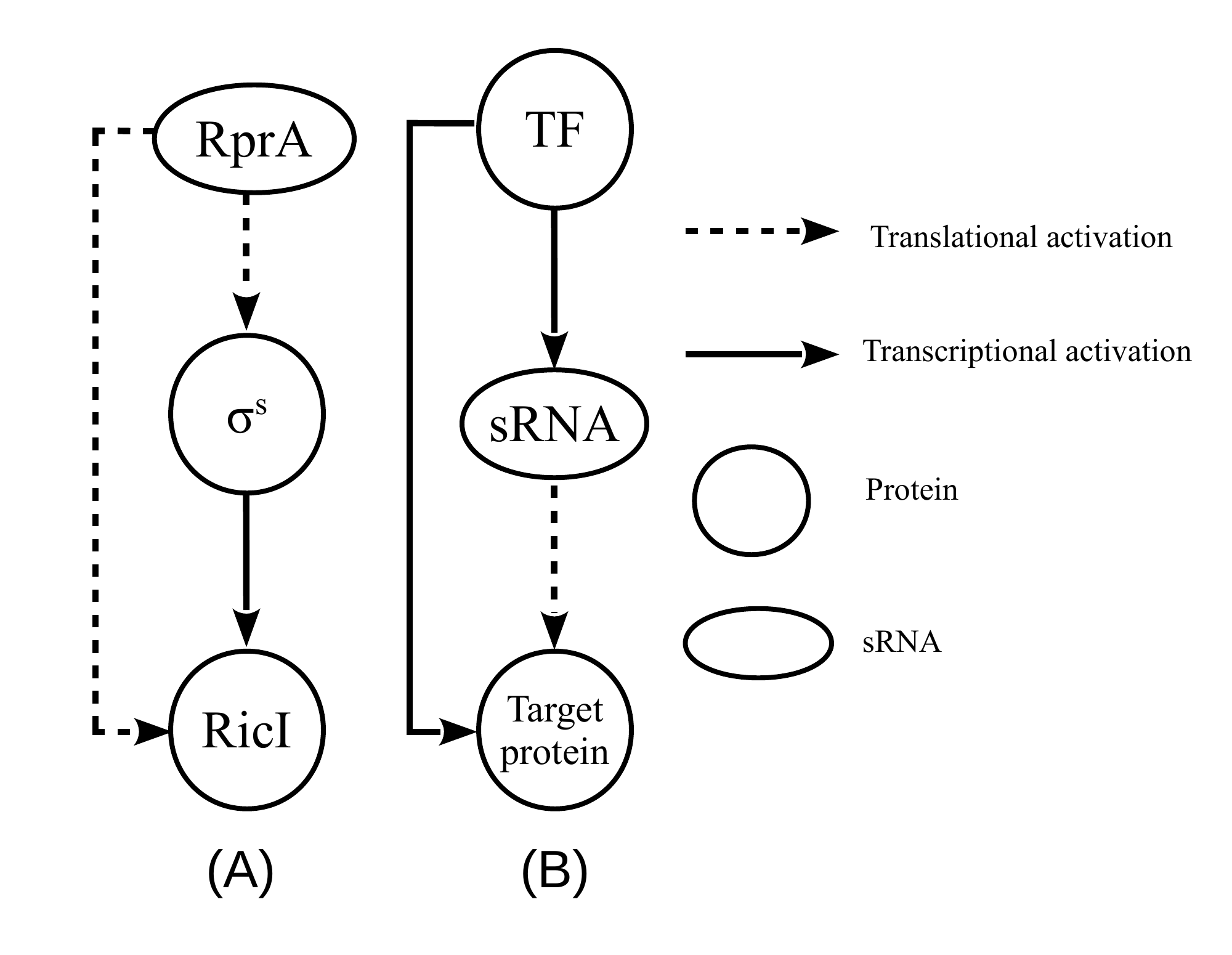}
 	\includegraphics[width=0.4 \textwidth]{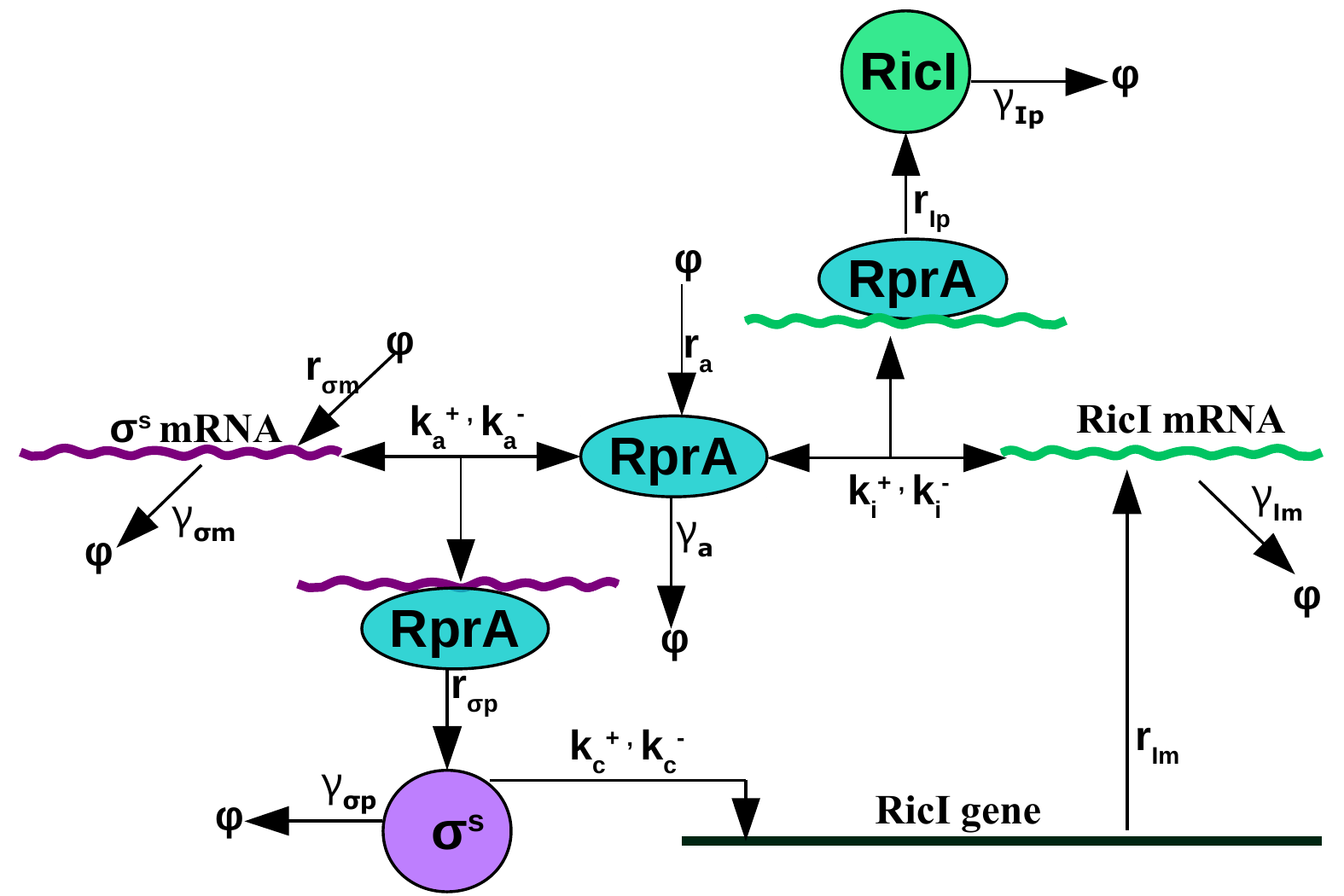}(C)
 	\caption{ (A) sRNA-driven coherent feed-forward loop  (sFFL). Here the upstream, master regulator is an sRNA, RprA.  (B) The sRNA-mediated 
 		coherent feed-forward loop (smFFL). Here the upstream, master regulator is a transcription factor (TF). The transcription factor activates the synthesis 
 		of the target protein mRNA  directly and also activates sRNA  synthesis  which in turn activates the translation of the target protein. 
 		The ellipse and the circle  represent sRNA and protein molecules, respectively. (C) A detailed diagram of sFFL
 		with various processes such as synthesis, degradation, etc. shown explicitly. $k^+$ and $k^-$, in general, represent the association and  dissociation 
 		rates, respectively, of protein regulators or sRNAs with the respective target gene or the target mRNA.  }
 	\label{fig:network-diagram}
 \end{figure*}

The process of  gene regulation is, in general,  non-linear  and involves  complex networks consisting of a number of genes, 
proteins, and sRNAs, which  are themselves   extensively regulated by  proteins, sRNAs etc.  
Despite the complexity, it is  possible to break down these complex regulatory   networks into smaller sub-networks that  function 
as basic building blocks of the bigger network.  
The sub-networks, typically known as network motifs, can then be analysed and based on the dynamics of individual motifs one may attempt to 
understand the dynamics of the entire complex network. The network motifs often  have recurrent occurrences and 
  typically certain specific types of network motifs, 
such as feed-back loops, feed-forward loops (FFL),  are over-represented. 
It is  believed that  these frequently occurring sub-networks  are naturally  chosen over others as they provide distinct  evolutionary 
 advantages such as  speeding up the response time, dampening of noise, etc.{\cite{alon1,mangan,shen}}.  
FFLs involving transcriptional regulators have been studied extensively  in the past {\cite{alon2}}. 
In a {\underline{t}}ranscriptional FFL (referred as tFFL from now on), a  protein 
regulator X  transcriptionally activates protein Z directly and also  indirectly through  the transcriptional activation
of an intermediate protein regulator Y which in turn activates Z transcriptionally. 
An FFL  where X positively (or negatively) regulates Z both directly and indirectly, is known as the coherent FFL. 
FFLs where the direct and indirect paths have opposite regulatory effects are referred as incoherent FFLs. 
Further, coherent (or incoherent) FFLs can be of different types depending on the nature of interactions (activating or inhibiting) in the individual paths \cite{mangan}.
FFLs involving purely transcriptional regulation are found, for example, in L-arabinose utilisation system \cite{schleif}. 
In addition to tFFL, there are other types of FFLs where Y is an 
sRNA  regulating the translation or the stability  of the target mRNA (of Z). 
Such FFLs will be referred in the following as \underline{s}RNA-\underline{m}ediated FFL (smFFL) from hereon. 
Another interesting variation is an FFL, wherein the upstream regulator (for example, X,  as introduced above) is an sRNA that drives the entire 
FFL. sRNA, in general, may activate or repress translation or the stability of downstream mRNAs. 
We shall refer to such FFL  in the following  as \underline{s}RNA-driven feed-forward loop (sFFL).  The comparison 
 between the two variants, sFFL and smFFL, is presented in 
 figures (\ref{fig:network-diagram}A) and (\ref{fig:network-diagram}B), respectively.
Through mathematical and computational modelling, a significant progress has been made 
in understanding  the role of FFL in  gene regulation. For example, mathematical modelling suggests that a coherent  tFFL can 
reject transient activation in X and respond only to persistent activation  of X {\cite{mangan}}.  
Similar studies also elucidate diverse beneficial features of smFFL 
like a quick to response to sudden changes in the input signal  which is advantageous under transient stress conditions \cite{shimoni}, 
 the ability to dampen the fluctuations under different contexts \cite{osella,caselle,caselle2,samsonov,marinari} etc..
In this work, using different types of modelling schemes, 
we study  two different aspects of the sFFL;  one associated with the  
response of the system under different types of input signal, 
the other associated with the noise characteristics  of the network.

 The sFFL of current  interest  has a significant role in 
  horizontal gene transfer in {\it Salmonella enterica} {\cite{papenfort2}}. 
As figure (\ref{fig:network-diagram}A) and (\ref{fig:network-diagram}C) show, the FFL is driven by the sRNA, RprA, which 
  is one of the three sRNAs (others being  DsrA and ArcZ) that activate the translation of $\sigma^s$ mRNA that codes for the alternative  sigma-factor, $\sigma^s$ {\cite{gottesman3,gottesman4}}. 
    In this FFL, RprA sRNA drives the regulation in the synthesis of RicI protein through two parallel pathways, both involving  translational activation 
  by RprA.  RprA binds to $\sigma^s$  mRNA through base-pairing  and opens up a translation-inhibitory structure in the $5'$-untranslated region (5'-UTR) of the mRNA  to 
  	facilitate  ribosome binding and thereby promote $\sigma^s$ translation \cite{papenfort2}.
  $\sigma^s$ being a  transcriptional activator for RicI mRNA up-regulates RicI mRNA synthesis. Finally, RprA, by opening the translation-inhibitory 
  structure of  its other target, RicI mRNA, leads to an increased synthesis of RicI protein.  
    Thus, the  FFL involves  an  AND gate mechanism where both $\sigma^s$ and RprA sRNA 
    are essential for the  up-regulation of  RicI protein. 
    Different types of stress conditions are expected to activate $\sigma^s$ production. The AND gate mechanism, through 
       the involvement of both RprA and $\sigma^s$,  ensures that not every stress condition activating $\sigma^s$ synthesis 
   leads to over-expression of RicI protein.  

    The inter-cellular transmission of plasmid  through bacterial 
    conjugation has an important role in microbial evolution and survival {\cite{alberts}}. 
    It  is believed that this RprA driven
     FFL 
     is crucial for regulation in the   transfer of plasmid pSLT which encodes several virulence genes in {\it Salmonella}. 
    Plasmid transfer via bacterial conjugation is a complex, energy-intensive 
    process consisting of several proteins and  RNAs functioning in concert. An important part of this process is the
     formation of pilus which establishes a  
     contact between the donor and the receptor cells and enables the transfer of genetic material. 
      Although such processes are found to be beneficial 
     in terms of adaptation in the changing environment,  
     due to significant fitness cost, such processes are  usually  tightly  regulated inside the cells {\cite{papenfort3}}.   
     It has been experimentally found that 
     the synthesis of  RicI protein is up-regulated  in {\it Salmonella} treated with   bile salt  which is a bactericidal agent. 
     Bile salt disrupts the bacterial cell membrane and under such membrane damaging activities, RicI protein interferes 
     with transmembrane assembly that, under a normal condition,  leads to  pilus formation. Thus, RicI protein   
       provides  an extra protection to the bacterial cell by 
   inhibiting energy expensive processes associated with pSLT transfer {\cite{papenfort2}}.
     
One of the major aims of mathematical modelling is to understand, in general, how a  network  processes or differentiates  different types   input signals such as 
sustained signal,  transient signal or oscillatory signal. 
These studies reveal, for example, the response or shut down time scales of the network under different types of input signals \cite{ mangan} 
or  sometimes more drastic response of the network depending 
on the details  of the input signal \cite{lingchong}.    Such temporal behaviour can be found out by solving the differential 
equations that  describe  how the densities of various regulatory components change with time.  The 
 differential equation framework essentially    captures  the  time variation of  the  average concentrations of various 
regulatory molecules. The gene expression is, however,   inherently noisy due to the probabilistic nature of various 
biochemical processes involved and the noise may lead to significant fluctuations in concentrations of the regulatory molecules \cite{raj}. 
 These fluctuations are especially relevant when the number of regulatory molecules is small in which case the noise may cause significant deviations 
 in the number of molecules  from the respective  average values. 
 An important  question in this case is that how, despite the inherent stochasticity, 
 the network controls the target protein level 
 reliably. Interestingly, it has been found that  there are specific network motifs such as negative feed-back 
loop or coherent transcriptional FFL that can filter or dampen fluctuations in comparison to many other network 
motifs \cite{alon1,bose1}. In this context, one of the pertinent questions is how the gene expression noise contributes 
to the target protein fluctuation in sFFL.

The aim of the present work is two-fold. The first objective is to use the differential equation description to 
quantitatively understand how the response of sFFL differs from the other types of FFLs, such as tFFL or smFFL. 
Interestingly, we find that the response of sFFL to an input signal is, in general, strong and 
rapid compared to tFFL or smFFL. Furthermore, it is possible to find approximate mathematical solutions for various concentrations 
by solving a system of coupled, non-linear equations under specific types of input signals. We show that the mathematical solutions 
agree reasonably well with exact numerical solutions of the differential equations. Such  mathematical solutions describe how the concentrations 
depend on various interactions, thereby reducing the need of further analysis on 
parameter variation. The second objective of the work is to explore the noise processing 
 characteristics of the sFFL and identify interactions that might be crucial in minimizing fluctuations in the  target protein concentration.
 We address this question by solving the stochastic model analytically using the master equation 
 based generating function approach \cite{vankampen}. This analysis leads to an interesting observation that the 
 present loop can effectively filter the noise and this noise filtering ability 
 depends  crucially on the synthesis rate of the upstream sRNA regulator and the degradation rate of the transcriptional activator 
 of the target protein.  In particular, we find a range of values of these parameters over which the noise attenuation becomes optimal.
 We verify our predictions through stochastic simulations based on Gillespie algorithm (GA), which yields 
 exact results on fluctuations in target protein concentration \cite{gillespie1,gillespie}. The simulation results agree well with mathematical predictions. 
 These observations prompt us to conclude that the master regulator sRNA, not only leads to strong and rapid response through 
 target protein synthesis, but it also plays a significant role in minimizing fluctuations in target protein concentrations.
 
\section{Results and discussions}
\subsection{Model and the deterministic description}    

The sFFL   of our interest  is  shown in figure (\ref{fig:network-diagram}A).  This can be compared with the 
smFFL  in figure   (\ref{fig:network-diagram}B). 
The change in 
concentrations of various regulators such as proteins, mRNAs and sRNAs  in sFFL 
with time is  described through the following differential equations
\begin{eqnarray}
&&\frac{d}{dt}[Ra]=r_a-\gamma_a [Ra], \ \ 
\frac{d}{dt}{[\sigma^sm]}=r_{\sigma m}-\gamma_{\sigma m}[\sigma^sm],\ \ 
\frac{d}{dt}[\sigma^sp]=r_{\sigma p} [Ra][\sigma^sm]-\gamma_{\sigma p} [\sigma^sp],\label{sigmam2}\\
&& \frac{d}{dt}[Rim]=f([\sigma^sp])-\gamma_{im} [Rim],\ \ 
 \frac{d}{dt}[Rip]=r_{ip} [Ra][Rim]-\gamma_{ip}  [Rip] \label{ricip2}.
\end{eqnarray}The notations used for various concentrations in these  equations are displayed in  table \ref{table-notation}. In general,  the synthesis  and degradation rates 
 are denoted by $r$ and $\gamma$, respectively.  Here, $f([\sigma^sp])=\frac{r_{im}[\sigma^s p]}{1+k_c[\sigma^sp]}$  is the Hill function representing transcriptional 
 activation in the synthesis of RicI mRNA by $\sigma^s$ protein and   $k_c$ is the ratio of activation and deactivation rates of RicI gene. $1/k_c$, 
 also known as the activation threshold, corresponds to the special  value of $\sigma^s$ protein concentration at which the transcription rate is the same as 
 $1/2$ of its maximum value. 
 \begin{table}[ht]
\caption{Notations used in the text}
\centering
\begin{tabular} {|c| c| c|}
\hline\hline
Protein or  mRNA or sRNA & Short form used in the text & Mathematical notation for concentrations  \\ \hline \hline
RprA sRNA & RprA & [Ra]  \\ 
$\sigma^s$ mRNA & $\sigma^sm$ &$ [\sigma^s m]$   \\ 
$\sigma^s$ protein  & $\sigma^sp$  & $[\sigma^sp]$ \\ 
RicI mRNA & RicIm  & $[Rim]$ \\ 
RicI protein & RicIp & $[Rip]$\\ \hline \hline
\end{tabular}\label{table-notation}\\
\end{table}

\subsection{Approach to the steady state}
\begin{widetext}
\begin{figure*}[ht]
 \includegraphics[width=0.4\textwidth]{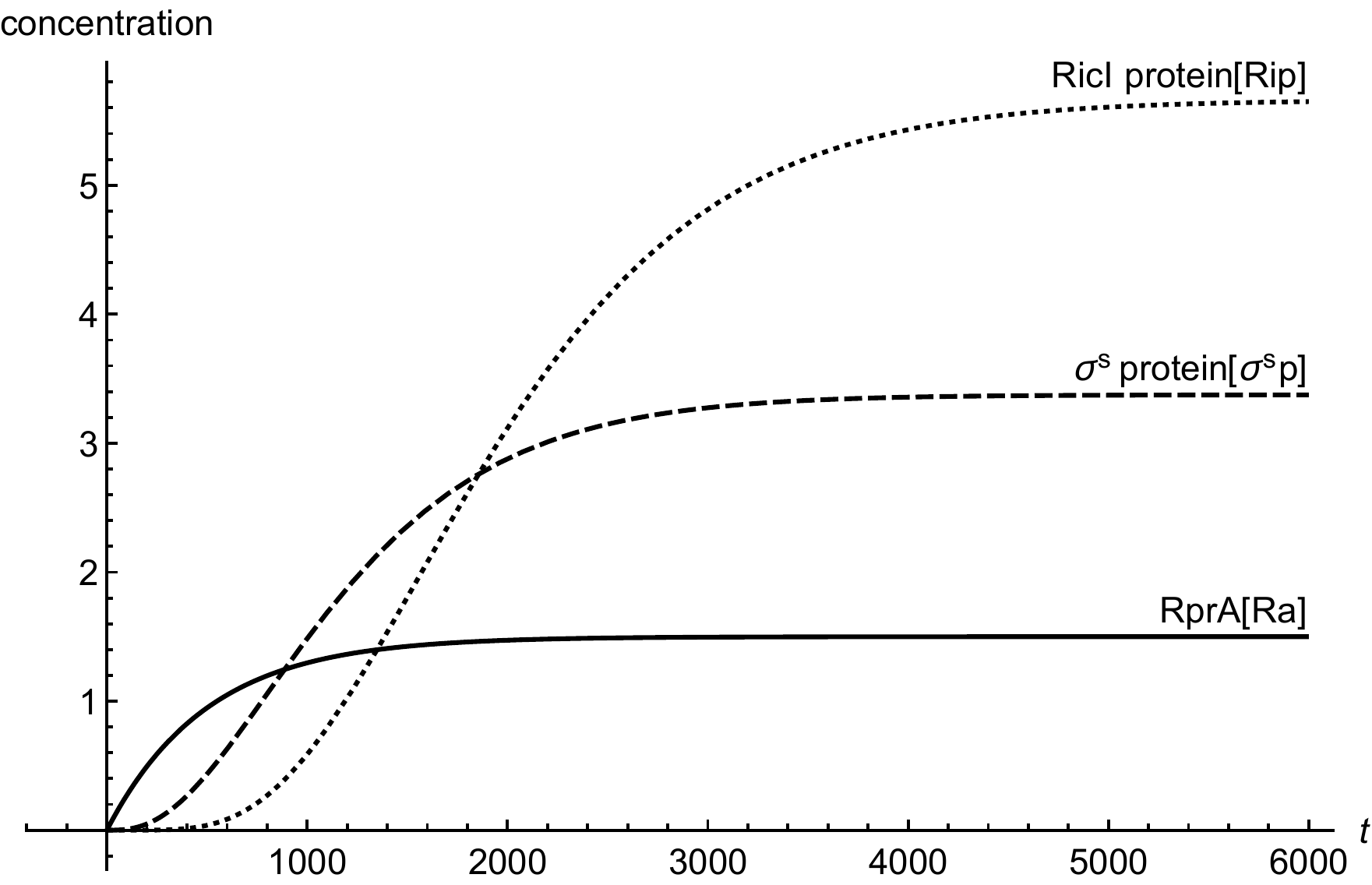} 
 \caption{ The  change in concentrations with time as the system approaches the steady state.  All synthesis and degradation rates are chosen 
 as $0.003$ (molecules. s$^{-1}$) and $0.002$ (${\rm s}^{-1}$), respectively and $k_c=0.3$ (${\rm molecule}^{-1}$).}
\label{fig:approach} 
\end{figure*}
\end{widetext}

The steady-state concentrations of various regulators can be obtained  by equating the time derivatives of various concentrations in equations (\ref{sigmam2}) and (\ref{ricip2}) 
to  zero and solving the  resulting algebraic equations. Figure (\ref{fig:approach})  shows  a comparison as 
 how different concentrations change with time  as they approach  the respective steady-state values.  For enabling a meaningful  comparison, 
 we have chosen the same synthesis  and degradation rates for all the regulatory components \cite{shen}. Unless mentioned otherwise, 
 for all the figures in the following, we assume negligible initial concentrations for all the regulators. 
  As the figure shows, the  increase in RicI protein concentration is slower compared to that of $\sigma^s$ protein initially (see also \cite{mangan,papenfort2}).  
  This lag in RicI production is due to the fact that the transcriptional activation of  RicI gene requires  production of sufficient amount of  transcriptional activator,
   $\sigma^s$.  The amount of delay in RicI production is, of course,  strongly dependent on  various biochemical parameters such as $k_c$ (the 
   activation threshold for  RicI mRNA transcription), degradation and synthesis rates of various regulatory components.  

\subsection{Temporal Solution} In order to understand the functionality of the present network motif in comparison with other motifs such as tFFL and smFFL, 
we solve the differential equations numerically for these three types of motifs.  The  differential equations describing the dynamics of the 
tFFL and smFFL are shown in appendix \ref{sec:app1}. For sFFL, we consider differential equations in (\ref{sigmam2}) and (\ref{ricip2}). 
Figures (\ref{fig:smallinitial}) and (\ref{fig:largeinitial}) show a  comparison as how the target protein concentration approaches the steady state  
with time in three different cases.  In order to have a  meaningful comparison, we consider the same parameter values for all the three motifs. While for  figure (\ref{fig:smallinitial}), the initial concentration of the upstream, master regulator is low, in case of figure  (\ref{fig:largeinitial}) the initial concentration of the upstream regulator is relatively high.  Irrespective of the initial condition, the response of the sFFL is  the fastest among all the three motifs.  The difference in the regulatory patterns seems to originate from the fundamental difference  in the mechanism of  transcritptional    and translational activations. The transcriptional activation that happens through binding to the DNA is expected to cause a delayed response since sufficient concentration of transcriptional activator is required for crossing the activation threshold.  In addition to this, due to the saturation kinetics, the transcriptional activation also reaches a saturation value  that is independent of the concentration of the transcriptional activator. This results in a delayed response in    tFFL that functions through  transcriptional activation at three different stages.   While in   smFFL, the  transcription interaction   is required for activation of two different genes,
 in sFFL, only   RicI gene is transcriptionally activated. It appears  that the fast response of sFFL is linked to a reduced number of transcriptional activation steps associated with the loop.  The difference due to the mode of regulation appears  more prominently in figure (\ref{fig:largeinitial}) where  an increased initial concentration in the upstream regulator leads to a  faster and stronger  response  in sFFL as compared to other FFLs.  
  The peak (non-monotonicity)  in the response curve, present in case of smFFL,  becomes more prominent in case of sFFL.  It might be that for  sFFL, the high initial   concentration of RprA and   the  direct interaction between  RprA and RicI mRNA   lead to a significant translational up-regulation in  RicI protein synthesis even  when the initial  RicI mRNA concentration is low.
 The sFFL  is activated  when the cell is exposed to  stress due to membrane-damaging activities of bile-salt. 
 RprA, along with $\sigma^s$ which activates RicI transcription, increases the expression of RicI protein. RicI  protein 
 localizes at the cytoplasmic membrane and, together  with other proteins, blocks the conjugation machinery required for pSLT transfer thereby providing protection to the cell by inhibiting energy-expensive conjugation  process. The rapid increase in RicI concentration especially when the initial RprA concentration is high,  is possibly  crucial for  cell's survival   under stress due to bactericidal agents such as bile salt.  
 
  \begin{widetext}
\begin{figure}
\begin{minipage}[b]{0.4\textwidth}
 \includegraphics[width=\textwidth]{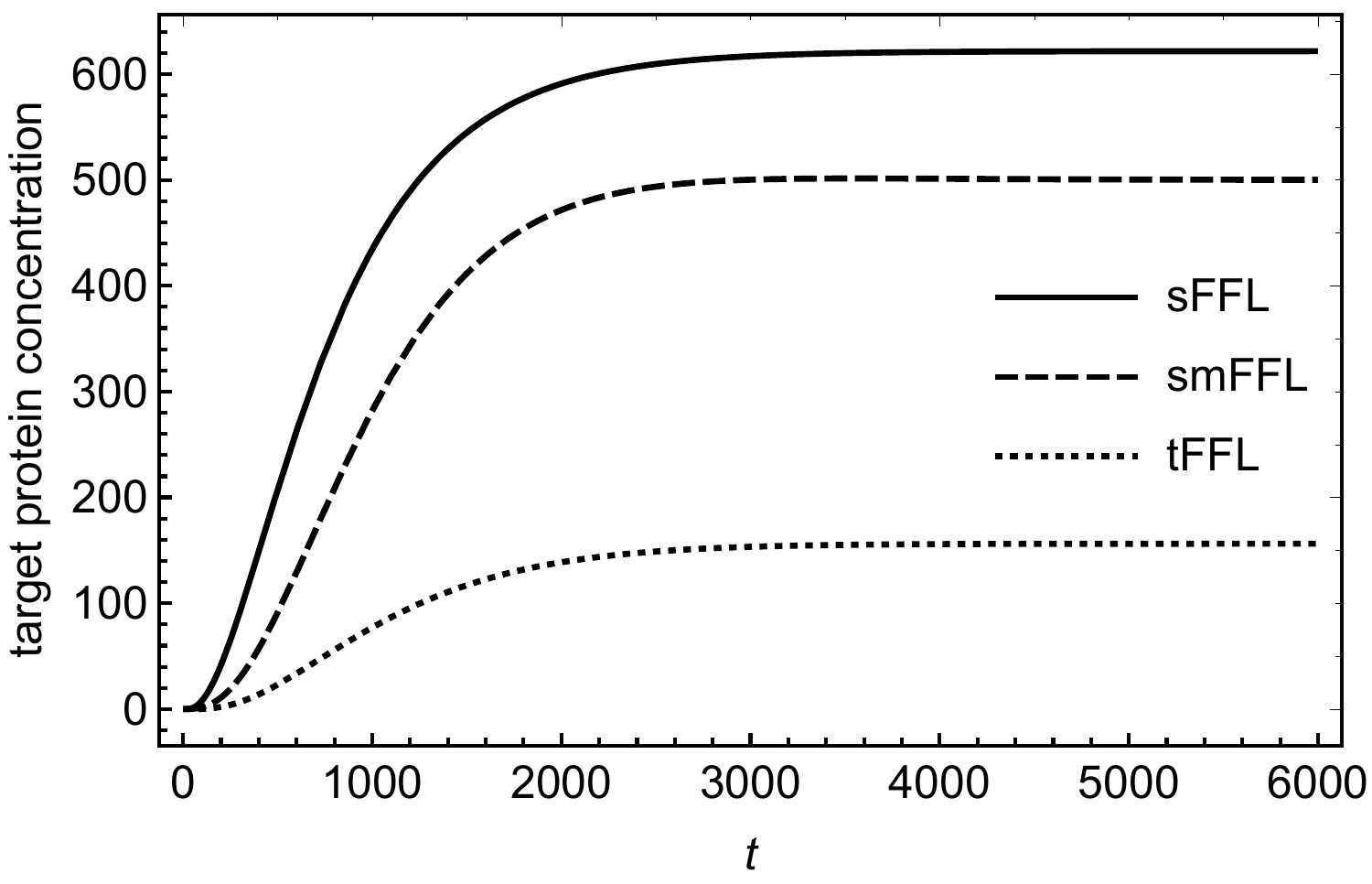} (A)
\end{minipage}
\begin{minipage}[b]{0.4\textwidth}
 \includegraphics[width=\textwidth]{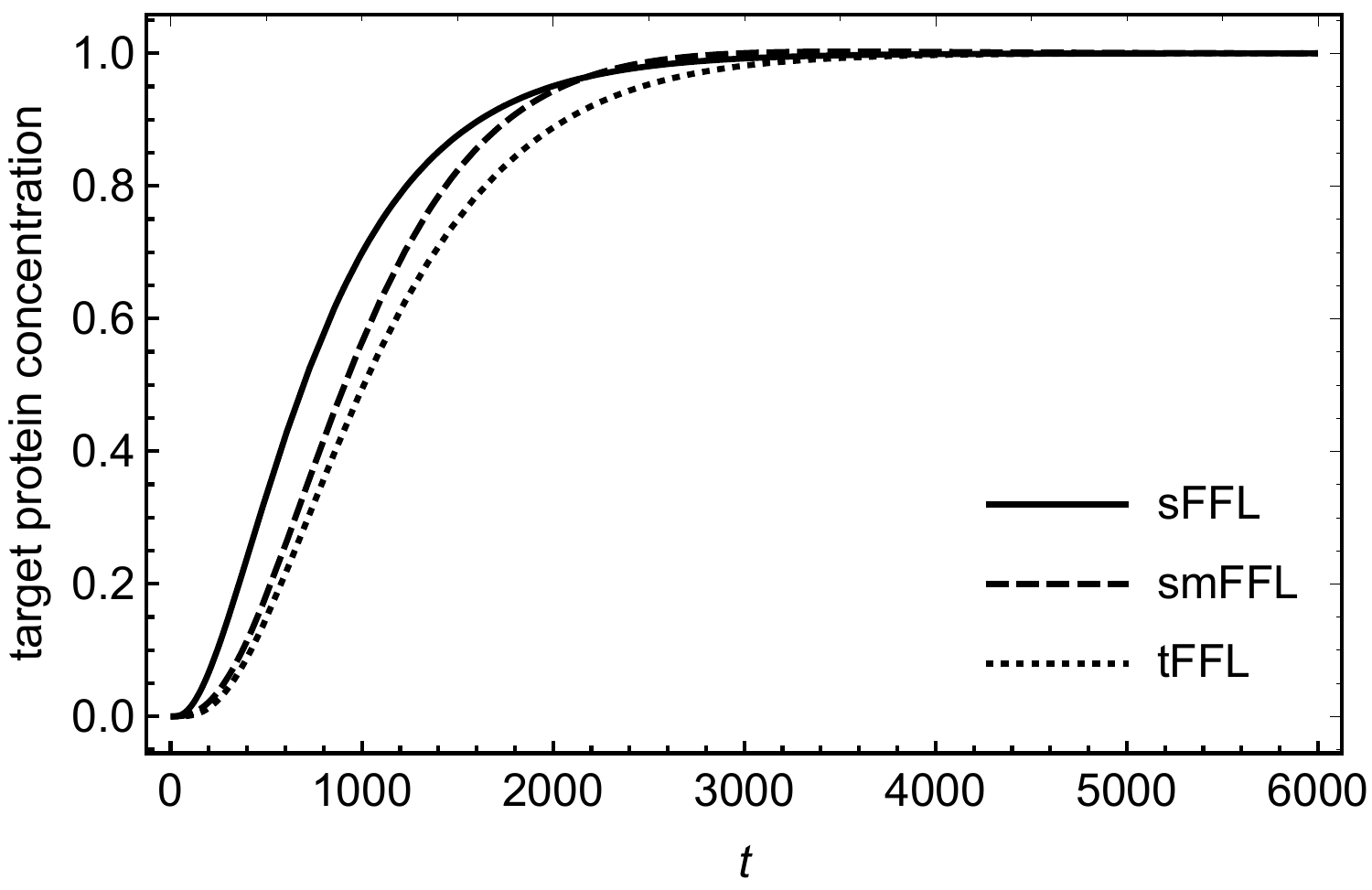} (B)
 \end{minipage}
 \caption{(A) The  change in the target protein concentration with time as the system approaches the steady state for tFFL, smFFl, and sFFL.  The  initial number of the upstream regulator 
 molecules is  $10$. 
 (B) The same plot with the concentration normalised by the respective steady-state values. To see the differences clearly, we have chosen the same synthesis rates, 
 $0.01$ (${\rm  molecules.\ s^{-1}}$) and degradation 
 rates  $0.002$ (${\rm s}^{-1}$) for all the regulatory molecules. Further, for sFFL, $k_c=0.3$ (${\rm molecule}^{-1}$). For other FFLs, the parameters (denoted by $k$ with various subscripts; see appendix \ref{sec:app1}) related to the association and dissociation constants in the Hill function  are chosen as $0.3$ (${\rm molecule}^{-1}$).}
 \label{fig:smallinitial}
 \end{figure}
 \begin{figure*}[ht]
  \includegraphics[width=0.4\textwidth]{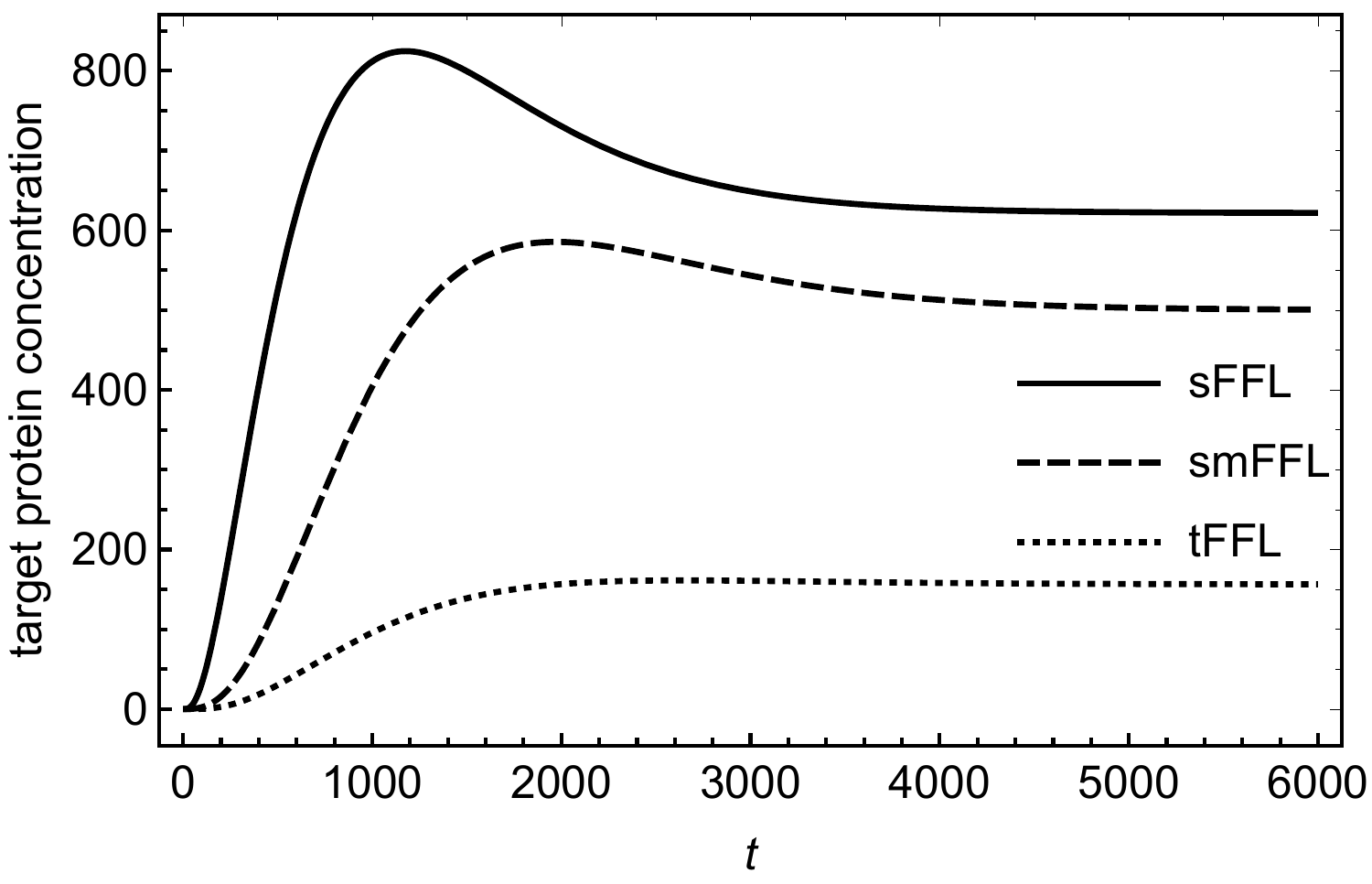} (A) 
  \includegraphics[width=0.4\textwidth]{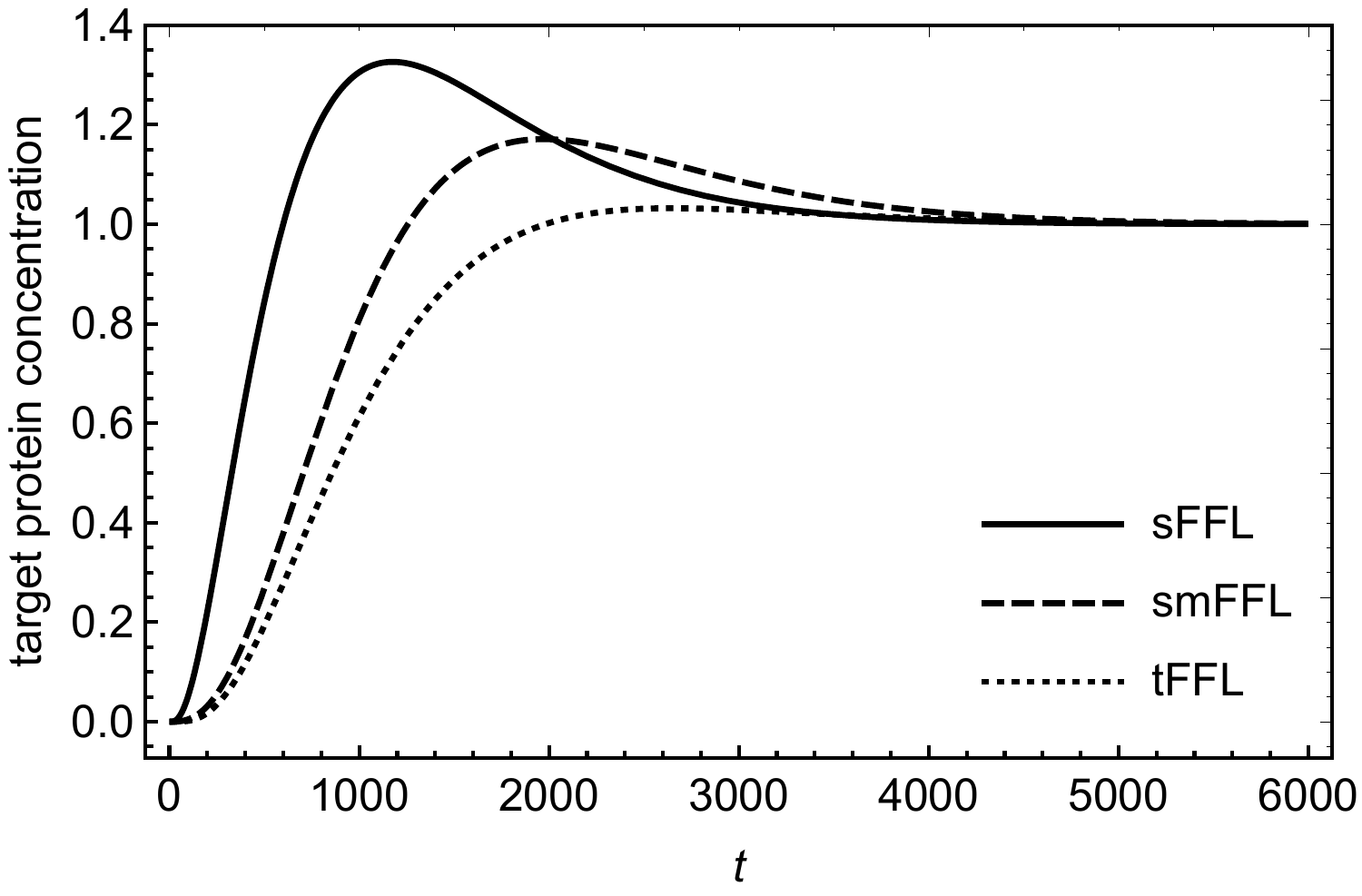} (B)
 \caption{(A) The  change in the target protein concentration with time as the system approaches the steady state for tFFL, smFFL, and sFFL. 
 The   initial concentration of the upstream regulator is   $30$.
 (B) The same plot with the concentration normalised by the respective steady-state values. Parameter values are same as those used for figure (\ref{fig:smallinitial}). }
\label{fig:largeinitial} 
\end{figure*}
\end{widetext}
We have, further,  obtained  approximate mathematical solutions (see appendix \ref{sec:app2} for details) 
for the target protein concentration in case of sFFL  for two different cases; (i) with a constant pool of 
upstream sRNA regulators and (ii) with sRNA  concentrations changing due to its synthesis and degradation processes. Mathematical solutions 
of such  coupled,  non-linear  equations can be obtained only over restricted parameter space. We show that 
 the mathematical solutions  agree reasonably well  with exact numerical  solutions of the differential equations in (\ref{sigmam2}) and (\ref{ricip2}) and  display  a 
  rapid growth in the target protein concentrations for large initial 
concentration of the upstream sRNA regulator as seen in figure (\ref{fig:largeinitial}).

\subsection{Noise processing characteristics}\label{subsec:generate}
\begin{figure*}[ht]
  \includegraphics[width=0.47\textwidth]{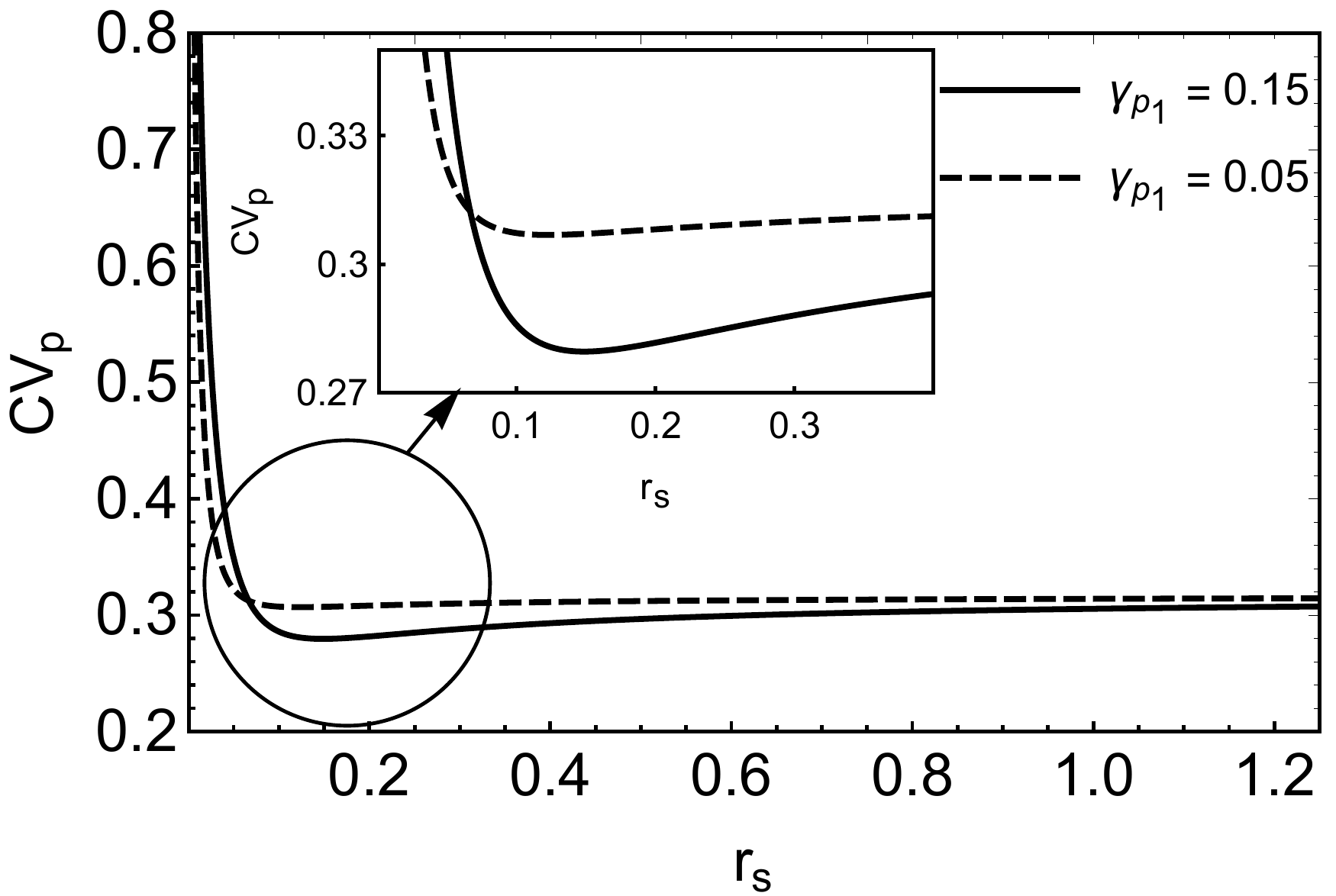}(A) 
  \includegraphics[width=0.47\textwidth]{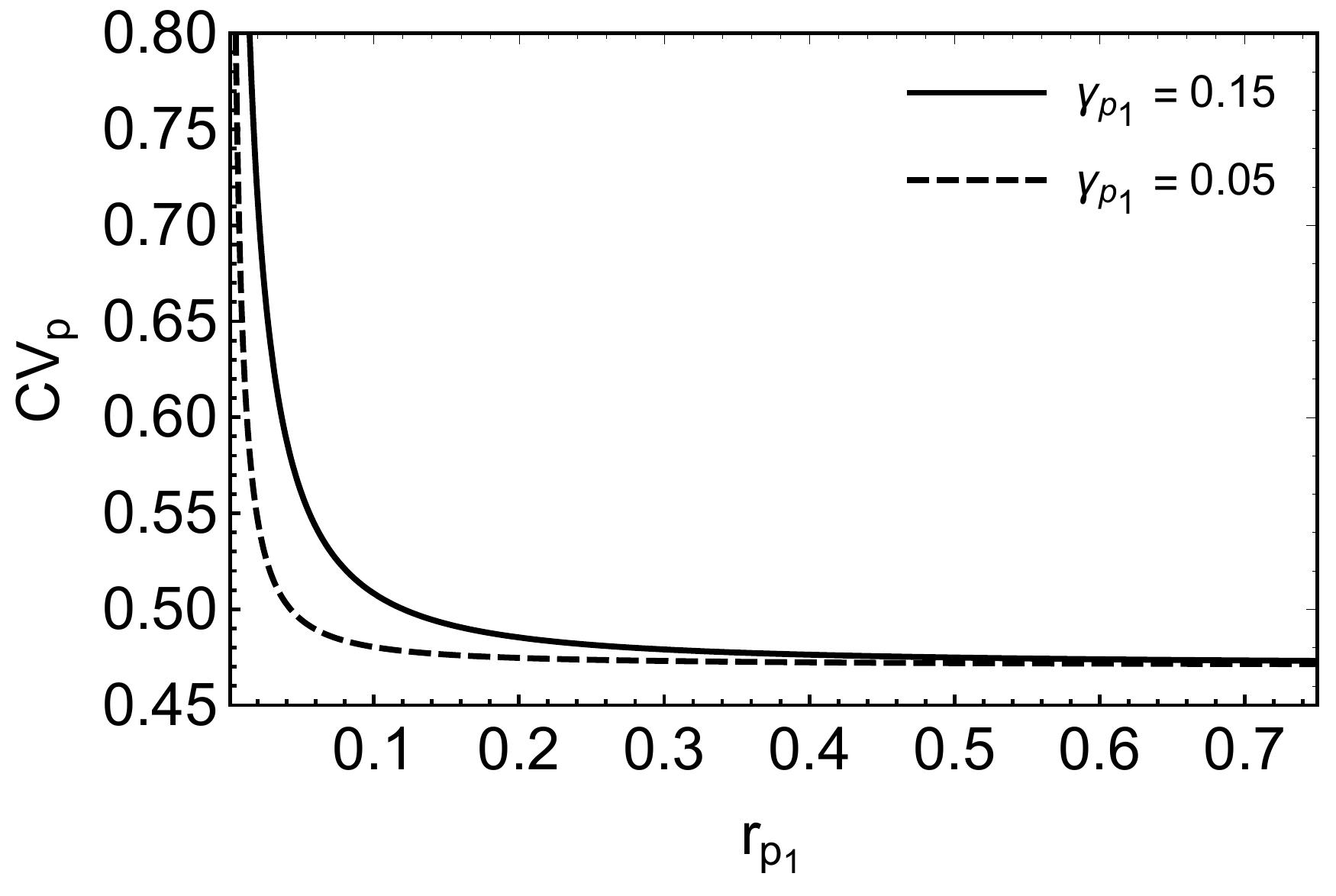}(B) 
 \caption{ Results from mathematical analysis. (A) The  coefficient of variation  for the target protein number  plotted with  $r_s$,  the 
 synthesis rate of sRNA, for different degradation rates, $\gamma_{p_1}$, of  protein, $p_1$.  
  The inset gives an enlarged view of the minimum region. (B) The  coefficient of variation  for the target protein number  plotted with $r_{p_1}$, the synthesis rate of the protein, $p_1$,  for different values of $\gamma_{p_1}$, the degradation rate of $p_1$. All the synthesis and degradation rates apart from those mentioned in the figure are 
  chosen as $0.01$ (${\rm  molecules.\ s^{-1}}$) and $0.002$ ($s^{-1}$), respectively. $k_c=0.1$ (${\rm molecule}^{-1}$).}
\label{fig:fluc} 
\end{figure*}

\begin{figure}
	\centering
	\includegraphics[width=0.7\linewidth, height=0.35\textheight]{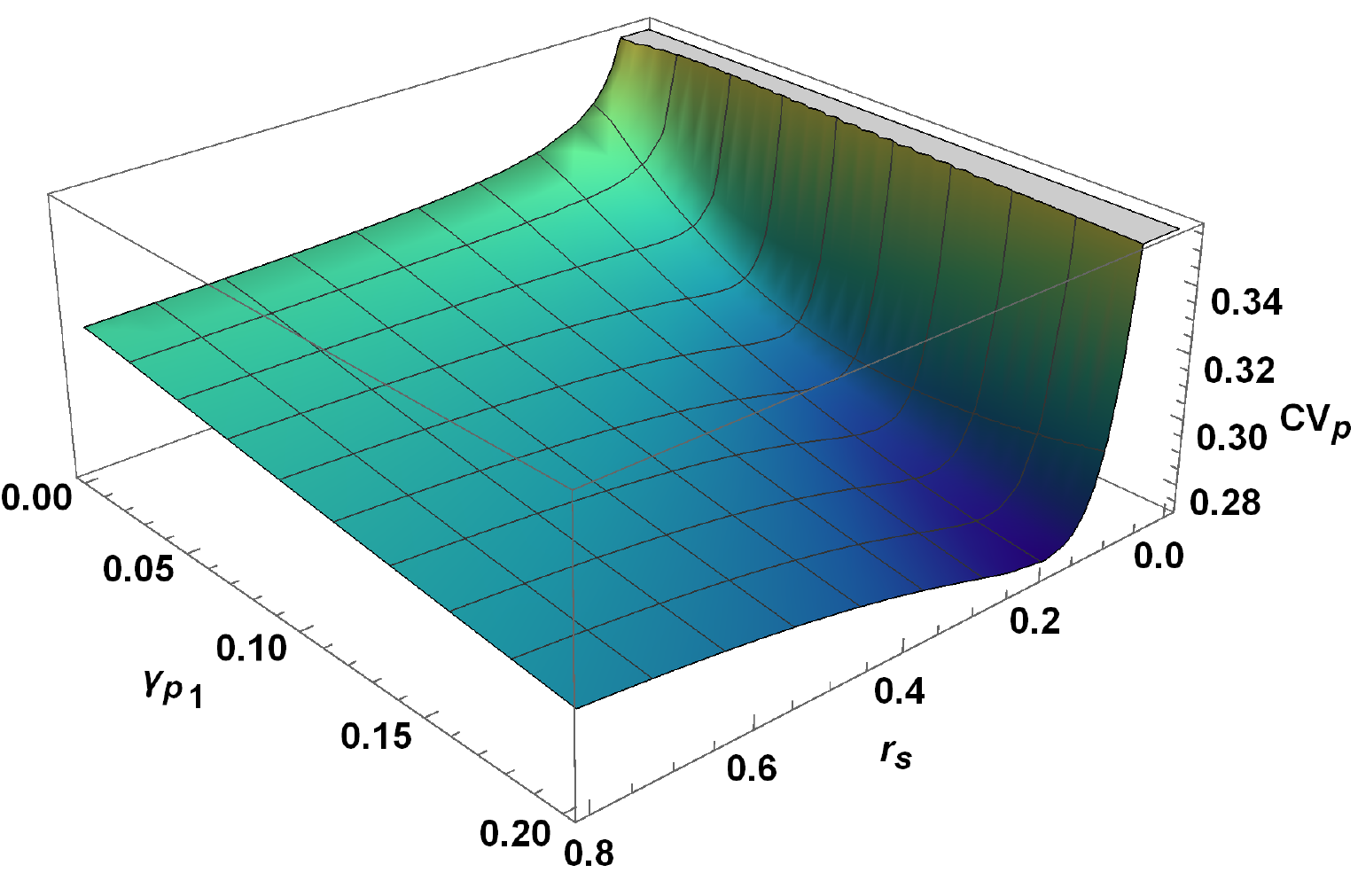}
	\includegraphics[width=0.1\linewidth, height=0.35\textheight]{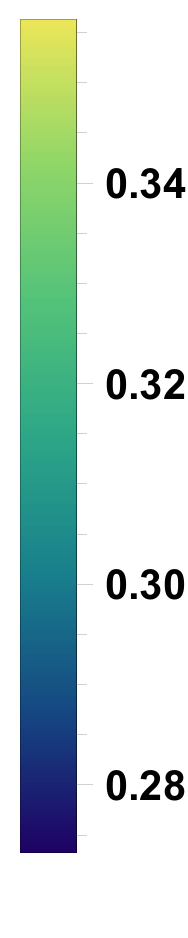}
	\caption{ A 3D plot of the coefficient of variation for the target protein with respect to $r_s$, the synthesis rate of the sRNA  
	and $\gamma_{p_1}$, the degradation rate of the intermediate protein, $p_1$.    The parameter values are as stated in figure (\ref{fig:fluc}). }
	\label{fig:3d-plot}
\end{figure}

Here we focus on  fluctuations in the target protein level in the sFFL of  present  interest.  
Using a general notation, we introduce  $s$, $m_1$, $m_2$, $p_1$ and $p_2$ as the number of sRNA (RprA), $\sigma^s$ mRNA, RicI mRNA, $\sigma^s$ protein and RicI  protein molecules, respectively.
This set of numbers represents a state of the system and $P_{s,m_1,m_2,p_1,p_2}(t)$ denotes the probability that the system is in  
a given state specified by these numbers at  time $t$. 
Our fluctuation analysis is based on the master equation which is a differential equation  that describes how this probability changes with time {\cite{osella}}.  The  probability changes with time due to various 
processes such as transcription, translation  and degradation of  different types of 
 molecules as per the details of the network. The  master equation that  takes into account all these processes  is 
\begin{eqnarray}
\frac{\partial}{\partial t} P_{s,m_1,m_2,p_1,p_2}&& =r_s (P_{{s-1},m_1,p_1,{m_2},p_2}-P_{s,m_1,p_1,{m_2},p_2}) + \gamma_s ((s+1)P_{s+1,m_1,p_1,{m_2},p_2} - s P_{s,m_1,p_1,{m_2},p_2})+\nonumber\\
&&\ r_{m_1} (P_{s,m_1-1, p_1,{m_2},p_2}-P_{s,m_1,p_1,{m_2},p_2}) + \gamma_{m_1} ((m_1+1)P_{s,m_1+1,p_1,{m_2},p_2} - m_1 P_{s,m_1,p_1,{m_2},p_2})+\nonumber\\
&&\ r_{p_1}\ s\ m_1  (P_{s,m_1,p_1-1,{m_2},p_2}-P_{s,m_1,p_1,{m_2},p_2})+ \gamma_{p_1} ((p_1+1)P_{s,m_1,p_1+1,{m_2},p_2} - p_1 P_{s,m_1,p_1,{m_2},p_2})+\nonumber\\
&&\ r_{m_2}(p_1) (P_{s,m_1,p_1,{m_2}-1,p_2}-P_{s,m_1,p_1,{m_2},p_2})+ \gamma_{m_2} (({m_2}+1)P_{s,m_1,p_1,{m_2}+1,p_2} - {m_2} P_{s,m_1,p_1,{m_2},p_2})+\nonumber \\ 
&&\ r_{p_2}\ s \ {m_2}\ (P_{s,m_1,p_1,{m_2},p_2}-P_{s,m_1,p_1,{m_2},p_2}) + \gamma_{p_2} ((p_2+1)P_{s,m_1,p_1,{m_2},p_2+1} - p_2 P_{s,m_1,p_1,{m_2},p_2})\label{prob-noise}
\end{eqnarray}
Here  $r$  and $\gamma$, in general, represent synthesis and degradation  rates, respectively, of different types of molecules. $r_{m_2}(p_1)$  denotes the transcription rate under 
transcriptional activation by $p_1$ and it is represented by the Hill function as $r_{m_2}(p_1) = \frac{r_{m_2} k_c\ p_1}{1 + k_c\ p_1}$, where $1/k_c$, as before, 
 denotes the activation threshold. In order to proceed further, we approximate the Hill function about the average density $\langle p_1\rangle$ at the steady state.  Under such approximation, we have 
\begin{eqnarray}
r_{m_2}(p_1)=r_{m_2}^0+r_{m_2}^1 p_1,\ \ {\rm where} \ \ r_{m_2}^0 = \frac{r_{m_2} k_c^2 \left\langle p_1\right\rangle ^2}{{(1 + k_c  \left\langle p_1\right\rangle) }^2}\qquad {\rm and} \qquad r_{m_2}^1 = \frac{r_{m_2} k_c}{{(1 + k_c  \left\langle p_1\right\rangle)}^2 }
\end{eqnarray}
Next, we  introduce the moment generating function 
\begin{eqnarray}
G(z_1,z_2,z_3,z_4,z_5) = \sum_{s,m_1,p_1,{m_2},p_2} z^s_1\ z^{m_1}_2\ z^{p_1}_3\ z^{m_2}_4\ z^{p_2}_5\ P_{s,m_1,p_1,{m_2},p_2}\label{eq:mgf}
\end{eqnarray}
with  $G\mid_{\{z_i\}=1}=1$. Various derivatives of the generating function  are related to average quantities as   
  $\frac{\partial}{\partial z_i}G(z_1,z_2,z_3,z_4,z_5)\mid_{\{z_i\}=1}=\langle n_i\rangle$ and  
$\frac{\partial^2}{\partial z_{i}^2}G(z_1,z_2,z_3,z_4,z_5)\mid_{\{z_i\}=1}=\langle n_i^2\rangle-\langle n_i\rangle$ where $\langle n_i\rangle$ denotes 
 the mean number of molecules of the $i$th species.  Denoting $G_{ij}$ as $G_{ij}=\frac{\partial^2}{\partial z_{i}\partial z_{j}}G\mid_{\{z_i\}=1}$, one may find the 
fluctuation in the target protein concentration as $G_{55}+G_5-G_5^2$.  
Using (\ref{prob-noise}), it can be seen that the  moment generating function satisfies the following equation
\begin{eqnarray}
\partial_t G=&&r_s (z_1-1)G+ \gamma_s (1-z_1)\partial_{z_1}G  +r_{m_1}(z_2-1)G+ \gamma_{m_1} (1-z_2)\partial_{z_2}G +r_{p_1}z_1z_2   (z_3-1)\partial^2_{z_1z_2}G +\gamma_{p_1} (1- z_3)\partial_{z_3}G
\nonumber\\
&&+\ r_{m_1}^0 (z_4 -1)G+ r_{m_1}^1 z_3 (z_4  -1) \partial_{z_3}G +r_{p_2}z_1 z_4(z_5 -1)\partial^2_{z_1 z_4}G + \gamma_{m_2} (1 -z_4)\partial_4 G+ \gamma_{p_2} (1 -z_5)\partial_5 G,\label{diff-generating}
\end{eqnarray}
where $\partial_{x}G=\frac{\partial}{\partial x}G$. In the steady state ($\partial_tG=0$), one may obtain successive moments by taking successive derivatives of equation (\ref{diff-generating}) and 
then  substituting $\{z_i\}=1$ in the resulting equations.  In order to obtain  the target protein fluctuations, we need  to find the first and second moments such as $G_{5}$ and $G_{55}$.  The   evaluation of these 
moments becomes complex since  an equation for   a moment, in general, involves the  higher order moments. In the present case, we are able to obtain 
 the target protein fluctuation by evaluating up to fourth moment and ignoring the contributions of the higher order moments (see  appendix \ref{sec:app3}
for details).  The coefficient of variation for the target protein, $CV_p=({\langle p_2^2\rangle-\langle p_2\rangle^2})^{1/2}/{\langle p_2\rangle}$,  
has been plotted in figure (\ref{fig:fluc}A). The figure shows a minimum indicating  an  optimal attenuation  of    
fluctuations in the target protein level  as  $r_s$ is changed. 
Similar optimal noise filtration can also be seen with respect to $\gamma_{p_1}$, the degradation rate of the transcriptional activator, $p_1$.  
Figure (\ref{fig:3d-plot}) shows a 3-dimensional picture of  optimum noise filtration as $r_s$ and $\gamma_{p_1}$ are changed. 
These figures show that  an  optimal 
noise attenuation is seen  only over an intermediate range of values of $r_s$ and $\gamma_{p_1}$ and the minimum disappears 
as one increases the values of these parameters. 
Further, with similar parameter values, no optimal filtration of noise can be found with respect to the synthesis rates of $m_1$ and $p_1$ 
(see, for example, figure (\ref{fig:fluc}B)). 

 For understanding  the rationale behind such noise processing characteristics of the sFFL, we compare these results with the case where 
 gene regulation involves  only translational activation by sRNAs (see figure (\ref{fig:srna-regulation}A)  in appendix (\ref{sec:app4})). 
 This is equivalent to considering only   the  direct pathway  in the sFFL through which RprA translationally 
 activates RicI mRNA. This is also how RprA up-regulates $\sigma^s$ translation in  the present sFFL 
 while $\sigma^s$ transcripts are synthesized  in the cell under various 
 stress conditions. Various moments for such a minimal motif  can be obtained  in a straightforward way 
 from the details presented in  appendix \ref{sec:app3} for the sFFL.  The coefficients of variation 
 for the target protein of  such  a minimal motif  and the sFFL are shown in figure (\ref{fig:srna-regulation}B) in appendix \ref{sec:app4}. The figure clearly 
  shows that the noise  processing characteristics  of the  simple motif with only  translational activation by sRNA  are significantly different from that of 
 sFFL considered here. In  particular,  no optimal noise filtration can be found in case of the minimal motif.  
 Further, for the parameter values chosen here, near the region of optimal 
 attenuation of noise, the  coefficient of variation of sFFL becomes  approximately $85\%$  of  that of the simple case  
 of gene regulation through translational activation. Despite having additional regulatory molecules and hence, associated fluctuations, the sFFL seems 
 to be less noisy as compared to the simple sRNA regulation scheme discussed here. Further,  it appears that the topology of the  sFFL and the non-linear 
 interactions play a crucial role in noise filtration characteristics.  

\begin{figure*}[ht]
\includegraphics[width=0.4\textwidth]{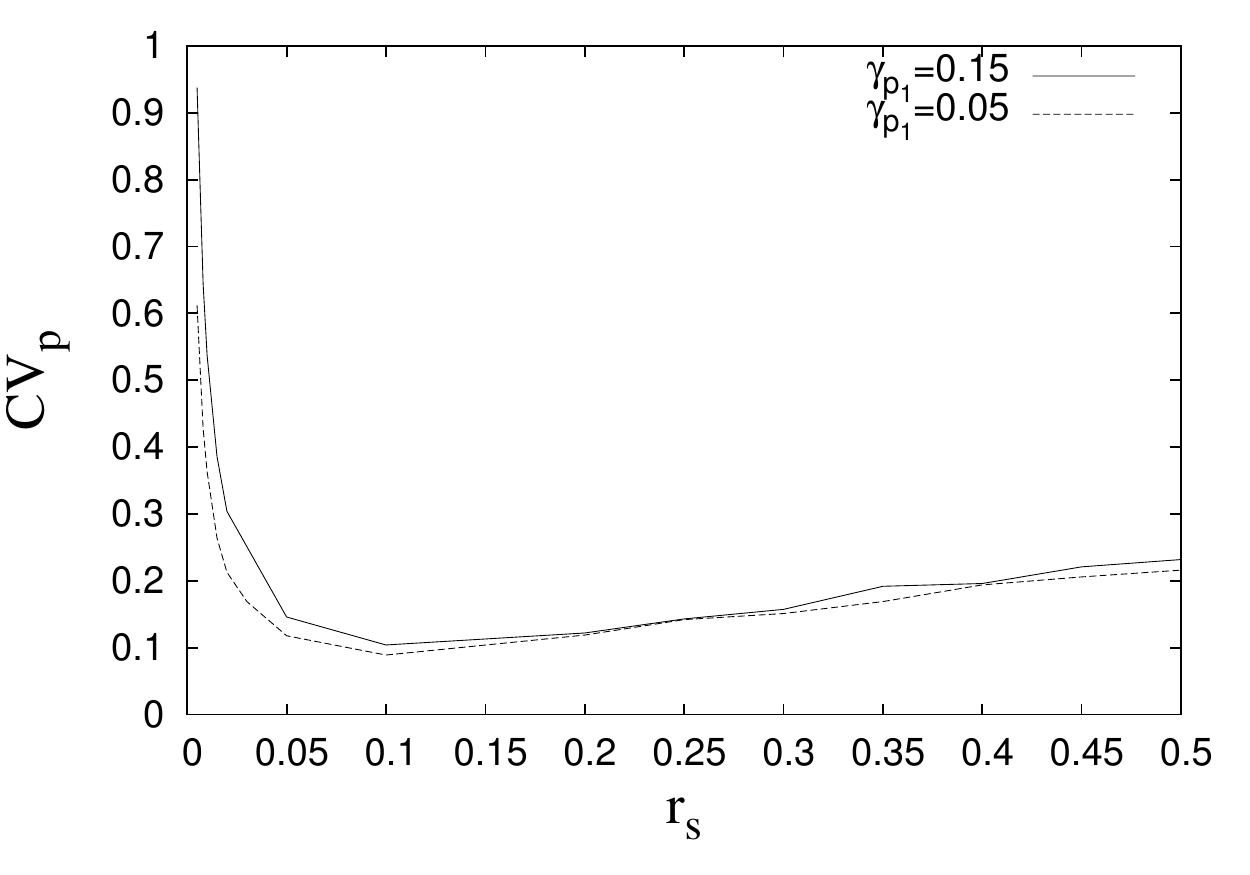} (A)
\includegraphics[width=0.4\textwidth]{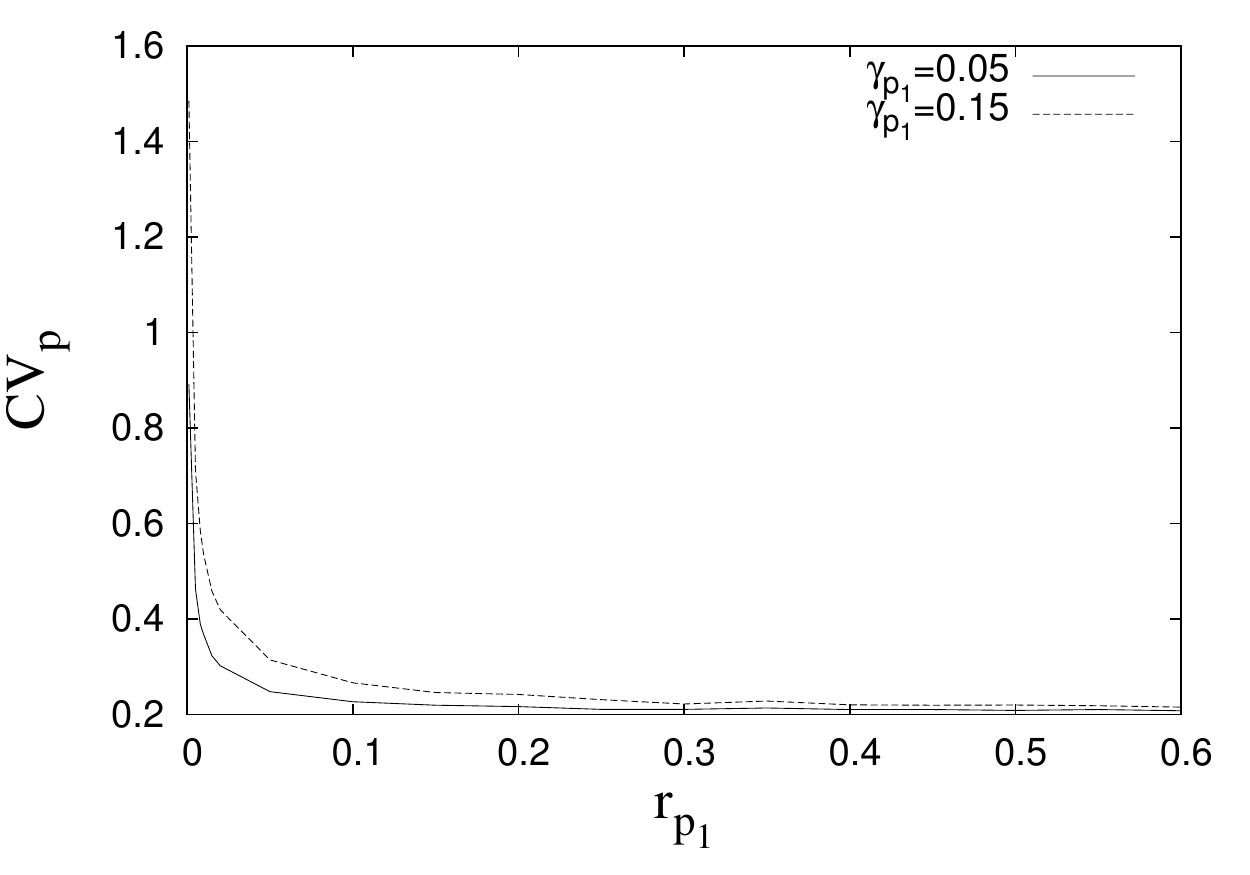} (B)
 \caption{ Results from stochastic simulations. (A) The  coefficient of variation  for the target protein number with the  synthesis rate of sRNA,  $r_s$, for different degradation rates of 
 $p_1$ protein, $\gamma_{p_1}$. 
    (B) The  coefficient of variation  for the target protein number with  synthesis rate of the $p_1$ protein, $r_{p_1}$, for different degradation rates of $p_1$ protein, 
    $\gamma_{p_1}$. }
\label{fig:fluca} 
\end{figure*}

The results from the above mathematical analysis are supported by stochastic simulations  discussed below. 
Stochastic simulations  based on Gillespie algorithm (GA)
 is an exact method that allows us  to incorporate 
the  probabilistic features in the gene expression in a systematic way {\cite{gillespie,gillespie1}}. 
In GA, we take into account  different types of   molecules  such as  protein, 
sRNA, mRNA  molecules  associated with the regulatory network and the relevant interactions between them. 
The number of each species of molecules at a given 
instant of time is denoted by $x_i(t)$ where $i=1,\ 2,\ 3,...N$ denotes   a specific species of molecules given that there are totally $N$ different species of molecules. 
GA is based on two  assumptions;  (1) the time 
interval between two successive reactions  is a random variable obeying Poisson Distribution  and (2) 
 the specific reaction that occurs at a given instant of time is selected randomly. 
In the simulation, this is executed  by  drawing  two random numbers at each simulation  step; 
one  random number, $r_1$, is used to  decide the time interval between two successive reactions and the other 
random number, $r_2$,  is used to  decide which reaction 
would occur in the next time step. Denoting a specific reaction by  $\mu$, where  $\mu=1, 2, 3 ...$, one may define a probability $a_\mu(t) dt$ which indicates the 
probability of occurrence of the $\mu$th reaction in the time interval $t$ and $t+dt$. Here $a_\mu(t)$ is expressed as 
the product of two factors; one is the reaction rate  and the 
other factor is the number of possible $\mu$ type reactions. Thus 
at   each simulation  step, the time interval before the next reaction  is determined and depending on the value of  $r_2$ and the various reaction 
rates $a_\mu(t)$, the next reaction is chosen.
Subsequently, based on which reaction has occurred, the  values of appropriate  $x_i$ are updated. 
We allow the system to evolve over $5\times10^7$ simulation  steps and keep a record 
 of the number of sRNA,  mRNA, target protein number 
  after  every $500$  simulation  steps leaving about $2\times 10^4$ initial  steps. 
     The details of the reactions considered and the corresponding rate constants are presented in  appendix (\ref{sec:app5}).
 The coefficients of variation for the target protein is shown in figures (\ref{fig:fluca}A) and (\ref{fig:fluca}B). 
 Consistent with the mathematical analysis  (figure (\ref{fig:fluc}A)), figure (\ref{fig:fluca}A) 
 also shows an optimal noise attenuation with $r_s$, the sRNA synthesis rate. 
 As   figure (\ref{fig:fluc}B), figure (\ref{fig:fluca}B)  shows no minimum in the coefficient of variation as $r_{p_1}$  i.e. the synthesis rate of protein $p_1$  is changed.  
  
  The noise processing characteristics of  different types of FFLs have been studied in the past. Using a Langevin description, it was shown that among various types of  tFFLs, the 
 coherent tFFL with all three activating regulatory interactions,  was the  least noisy  among other kinds of coherent tFFL \cite{bose1}. In particular, it was shown that 
 the activating nature of all  regulatory interactions is responsible for such maximal noise reduction.   Recently,  noise processing characteristics 
 have also been studied  in detail for incoherent micro RNA (miRNA) mediated FFL \cite{osella, caselle,caselle2, marinari}. Unlike Fig. 1B, in such miRNA mediated FFL, the miRNA 
 represses the target protein synthesis  by binding to the  mRNA of the target protein and 
 degrading it subsequently.   Since sRNA/miRNA mediated target repression is a common scenario in gene regulation,  there have  been efforts in the past 
 to understand the fundamental differences between the  miRNA/sRNA mediated and TF mediated gene repression \cite{hwa}. 
  Currently, it is  well established that, in general,  sRNA-mediated repression reduces   
 fluctuations in the target protein level as compared to TF mediated repression. Physically, this happens since in case of TF
 mediated repression, occasional transcriptional leakage is amplified due to translation causing  large fluctuations in the protein concentration.  In contrary to this, in case of   
 sRNA-mediated repression of the target mRNA, the mRNAs are rarely translated leading to a relatively smooth gene expression.  
 This feature of reducing fluctuations  in the target protein level  is also  found in incoherent  miRNA mediated FFL \cite{osella}.  Here, while the fluctuations in the 
 top-tier transcription 
 factor introduces similar variations  in both miRNA  and mRNA levels, the miRNA  due to its repression activity  reduces the target protein fluctuations against the transcription factor 
 fluctuations.  This leads to a significant noise attenuation in miRNA mediated FFL in comparison to gene expression regulated by transcription  factor alone. 
 Additionally,  it was also shown in \cite{osella} that the noise-filtration in  incoherent miRNA mediated FFL
   becomes optimal over a range of miRNA mediated repression strength, miRNA concentration and the  transcription factor concentration. 
 In contrary to the earlier work, the sFFL considered here has  all the regulatory interactions  activating in nature.  Interestingly, however,   this motif also shows   
  a significant noise attenuation at the target protein level  as compared to  a much simpler scenario where the  gene regulation involves only translational activation by sRNA.  
From this comparison,   it appears that   such noise processing behavior of the sFFL  is a result of   
a complex  interplay of the non-linear interactions  present in the two branches of the sFFL.  
It would be of considerable interest to explore if,  in the realistic scenario, the parameter values are tuned in a specific way  that  leads to 
maximum noise attenuation  in the target protein level. We believe that the mathematical  results for the fluctuations determined here 
can be used for exploring such possibilities.

\section{Conclusion}
In this paper, we have studied an sRNA-driven feed-forward loop (sFFL) through mathematical and computational modelling. Here the upstream, 
master regulator is an sRNA, RprA, which activates the synthesis of the final, target protein, RicI, using two parallel pathways; one, directly through 
translational activation of the RicI mRNA and the other, via translational activation of $\sigma^s$, which, in turn, functions as a transcriptional activator of RicI 
protein. Thus unlike other FFLs involving sRNAs as regulators at an intermediate level and proteins as the upstream regulator, here the upstream regulator 
is an sRNA while the intermediate regulator is a protein $\sigma^s$. This kind of an FFL was found only recently; and it is believed that such an FFL 
plays an important role in horizontal gene transfer in {\it S. enterica}. During horizontal gene transfer, bacterial conjugation happens through pilus formation, 
which is an energy-expensive process. 
In the presence of bactericidal agents, the RicI interferes with pilus formation and thereby inhibits horizontal gene transfer. This is one way to protect the 
bacteria during stress.  Hence, a quick response to stress necessitates a rapid increase in the levels of RicI. Our work presented here, allows us to rationalize 
how RprA driven FFL could be a productive strategy to achieve this control to rapidly interfere with pilus formation. In principle, this would be general to any 
sRNA-driven FFL.

While extensive studies to understand the general characteristics of FFLs, such as purely transcriptional FFL (tFFL) or sRNA-mediated feed-forward 
loop (smFFL) were carried out previously,  FFL driven by sRNA as the top-tier regulator, being a new variant, has not been subject of extensive investigations. 
To the best of our knowledge, the present work is the first, detailed modelling-based analysis of an sFFL. A major finding of this analysis is that an sFFL is 
capable of producing a strong and rapid response in terms of enhancing target protein levels,  compared to that by tFFL or by smFFL. By comparison, we 
believe, this would be a generic feature that is linked due  to the number of transcriptional activation steps present in sFFL as against tFFL or smFFL. While 
sFFL involves transcriptional activation of only one gene, for tFFL and smFFl two or more genes are transcriptionally activated. The saturation kinetics and the 
delay in response due to the presence of the activation threshold appear to be the reasons for delayed and weak response in case of transcriptional activation. 
This difference in the mode of regulation is captured by the mathematical equations for the respective FFLs and such generic features can be compared 
meaningfully, when the same parameter values are used for all three cases. A high sensitivity to the initial concentration of the upstream regulator, i.e. sRNA, 
and therefore a rapid initial response in the target protein level is an additional characteristic of sFFL. We have obtained explicit mathematical solutions describing 
how the concentrations of various regulatory molecules change with time. These solutions  
clearly demonstrate the sensitivity of the target protein concentration to the initial conditions. This phenomenon may  be rationalized as follows in the 
context of RprA driven inhibition of pilus formation in {\it S. enterica}. This sFFL is triggered during membrane-damaging activities of the bactericidal agent, bile salt, in 
order to arrest pilus formation by increasing RicI level rapidly. Hence, a sensitive and rapid response to the initial-conditions (concentration of sRNA) might be a 
necessary strategy for the cell to respond to stress. These insights can also be tested experimentally. 
A plasmid vector can be designed in a manner that the 
expression of RprA under the control of arabinose, can lead to the synthesis of $\sigma^s$ and RicI protein. In order to examine the levels of these proteins, 
reporter genes such as lacZ, GFP and RFP can be fused with the genes coding for $\sigma^s$ and RicI resulting in the synthesis of fluorescent labeled proteins. Such a design would allow a direct validation of the model proposed here.

Furthermore, the RprA driven FFL is expected to function reliably despite the presence of noise in  gene expression. This implies that the FFL should 
ideally filter out the noise such that the target protein (RicI) concentration does not fluctuate significantly from the average concentration level. 
Keeping this in mind, we analysed the noise characteristics of the sFFL through master equation based modelling and stochastic simulations. 
We find that the network indeed filters out the noise and this aspect depends significantly on the synthesis rate of the upstream 
regulatory sRNA and the degradation rate of the transcriptional activator of the target protein. More specifically,  an optimal attenuation of noise can be 
achieved  by varying the synthesis rate of sRNA or by varying the degradation rate of the transcriptional activator.  In order to  have a quantitative 
comparison of the 
noise filtering ability of the present loop with a simple sRNA-driven mode of regulation, we  consider   gene expression   regulated through translational activation  
by sRNAs alone. We find that, in this case, the  coefficient of variation in the target protein number  is significantly different from that of 
 sFFL and, in particular,   no optimal 
noise attenuation can be found. It appears that an optimal noise attenuation in sFFL is  a result of a complex interplay of the non-linear interactions present in the 
 two branches of sFFL. The results from this  stochastic analysis   can also be  verified experimentally by extending the  experimental design mentioned before 
suitably and quantitating the amount of mRNA and protein of RicI as a function 
of RprA concentration. 
Overall, the present work suggests that the function of sFFL  is critically governed 
by the sRNA - not only in generating a speedy and strong  response but also  in producing a reliable response by  regulating the  
gene expression noise. The prediction  about the  critical role of sRNA in noise filtering 
raises  new questions, such as, how the concentration of the top-tier sRNA is regulated in the cell in order to achieve optimal noise filtration through 
the FFL. Besides, this being a general model, the insights obtained from the present study will be applicable to other sRNA-driven coherent FFLs  
that might  be discovered in the future and also for designing artificial networks for optimal regulation. 

\bigskip
{\bf Conflict of interest:} Authors declare no conflict of interest. 

\bigskip
{\bf Acknowledgement}  ST and SM   thank DBT, India for financial support through grant no. BT/PR16861/BRB/10/1475/2016.
\section*{APPENDIX}
\setcounter{section}{1}
\appendix
\section{Purely transcriptional feed-forward loop and sRNA-mediated feed-forward loop}\label{sec:app1}

\begin{figure*}[ht]
	\includegraphics[width=0.1\textwidth]{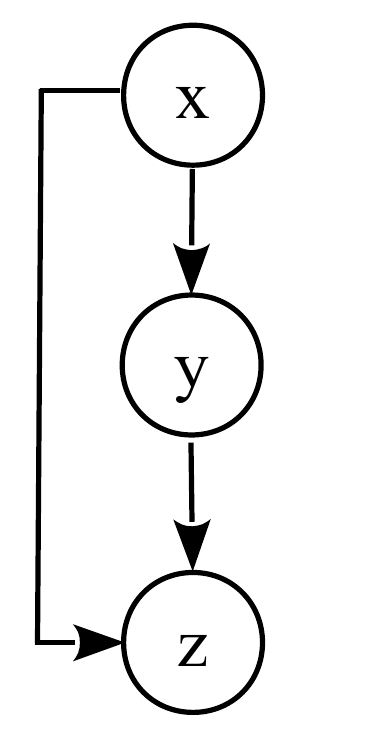} 
	\caption{A purely transcriptional feed forward-loop (tFFL). Circles represent  protein molecules. All the activation processes here are protein-mediated.}
	\label{fig:tf} 
\end{figure*}

In figures (\ref{fig:smallinitial}) and (\ref{fig:largeinitial}) of the main text, we have shown  how the   target protein concentration changes 
with time for tFFL and smFFLs. In the following, we present the differential equations that describe the dynamics of the tFFL and smFFL. A schematic representation of the
tFFL network is shown in figure (\ref{fig:tf}).
The parameters, $r$ and  $\gamma$, in general,  represent the  synthesis and degradation rates, respectively,  of various regulatory molecules. Parameters 
denoted by  $k$ in Hill functions are, in general,  related to the activation threshold. 

\subsection{Transcriptional feed-forward loop (tFFL)}
The following equations describe the tFFL.
\begin{eqnarray}
\frac{d}{dt}[x] &=& r_x -\gamma_x [x] \label{eqx},\\
\frac{d }{dt}[y_m] &=& \frac{r_{y_m} [x]}{1 + k_{xy} [x]} - \gamma_{y_m} [y_m],\\
\frac{d }{dt}[y_p] &=& r_{y_p} [y_m] - \gamma_{y_p} [y_p],\\
\frac{d}{dt}[z_m] &=& r_{z_m}\frac{ [x]}{(1 + k_{xz} [x])} \frac{[y_p]}{(1 + k_{yz} [y_p])}- \gamma_{z_m} [z_m] \ \ \ {\rm and}\\
\frac{d }{dt}[z_p] &=& r_{z_p} [z_m] - \gamma_{z_p} [z_p].\label{eqz}
\end{eqnarray}
$[x]$, $[y_m]$, $[y_p]$, $[z_m]$, and $[z_p]$ represent the  concentrations of the upstream transcriptional activator, $y$-mRNA, $y$-protein, $z$-mRNA, and $z$-protein, respectively. 
$x$ transcriptionally activates the synthesis of $y$-mRNA; $x$ and $y$ are both required for transcriptional activation of $z$. 
\subsection{sRNA-mediated feed-forward loop (smFFL)}
Equations describing smFFL are 
\begin{eqnarray}
\frac{d}{dt}[x] &=& r_{x} - \gamma_{x} [x] \label{smx},\\
\frac{d}{dt}[s] &=& \frac{r_s [x]}{1 + k_s [x]} - \gamma_s [s],\\
\frac{d}{dt}[m] &=& \frac{r_m [x]}{1 + k_m [x]} - \gamma_m [m],\ \ \ {\rm and}\\
\frac{d}{dt}[p] &=& r_p\ [m]\ [s] -\gamma_p [p].\label{smp}
\end{eqnarray}
Here, [x], [s], [m], and [p] denote the concentrations of   the upstream transcriptional activator, sRNA, target protein mRNA and target protein, respectively.  $x$ translationally activates sRNA 
synthesis as well as the synthesis of the target protein transcripts. 
Finally, the sRNA translationally activates the synthesis of the target protein, $p$.

\section{Details on Temporal Solutions }\label{sec:app2}
Here we find  the response of the network under  persistent  or  time-varying input signals. We assume two possible scenarios 
as a response to stress;  (a) a constant pool of sRNA, RprA,  (b) the concentration of sRNA, RprA, varies due to its synthesis and degradation. 
Under these conditions, we  explicitly find   the solutions of the differential equations  that describe how the concentrations of various regulators change 
with time. 

\subsection{Under a constant pool of sRNA}
Assuming a constant solution for RprA as $Ra(t)=R_0$,  and using a solution for $\sigma^sm(t)=c_{2S} e^{-\gamma_{\sigma m}t}+\frac{r_{\sigma m}}{\gamma_{\sigma m}}$, we find 
\begin{eqnarray}
[\sigma^s_p](t)=\frac{\sigma_{1S}}{\gamma_{\sigma p}-\gamma_{\sigma m}} e^{-\gamma_{\sigma m}t}+
\frac{\sigma_{2S}}{\gamma_{\sigma p}}+c_{3S} e^{-\gamma_{\sigma p}t},\label{eqsigmap}
\end{eqnarray}
where $c_{2S}$ and  $c_{3S}$ are  the integration constants. Considering negligible initial concentration of $\sigma^s$ mRNA and $\sigma^s$ protein, we find
$c_{2S}=-\frac{r_{\sigma m}}{\gamma_{\sigma m}}$ and $c_{3S}=-\frac{\sigma_{1S}}{\gamma_{\sigma p}-\gamma_{\sigma m}}-\frac{\sigma_{2S}}{\gamma_{\sigma p}}$.
Here $\sigma_{1S}=r_{\sigma p}  R_0 c_{2S}$  and $\sigma_{2S}=\frac{r_{\sigma p}  R_0 r_{\sigma m}}{\gamma_{\sigma m}}$. 
The concentration of RicI mRNA can be found in a similar way starting with $[Rim](t)=e^{-\gamma_{im}t} f_2(t)$.
The solution for $f_2(t)$ can be written as  
\begin{eqnarray}
f_2(t)=\frac{r_{im}}{k_c \gamma_{im}}e^{\gamma_{im}t}-\frac{r_{im}}{k_c\gamma_{im}} \int \frac{du}{A_S+B_S u^{-2}}+c_{4S},\label{eqricim1}
\end{eqnarray}
where $c_{4S}$ is the integration  constant, $A_S=1+\frac{k_c\sigma_{2S}}{\gamma_{\sigma p}}$ and 
$B_S=\frac{k_c\sigma_{1S}}{\gamma_{\sigma p}-\gamma_{\sigma m}}+k_c c_{3S}$. 
For the integration, we have used $u=e^{\gamma_{im}t}$. $B_S$ turns out to be negative 
upon substituting the expression for $c_{3S}$ in the definition of $B_S$. 
While  obtaining (\ref{eqricim1}), 
we have assumed that the degradation constant  of RicI mRNA, $\gamma_{im}$
is twice smaller than the degradation constants of $\sigma^s$ mRNA  and $\sigma^s$ protein $\gamma_{\sigma m}$ and $\gamma_{\sigma p}$, respectively 
and $\gamma_{\sigma p}\approx \gamma_{\sigma m}$. 
The final solution  for RicIm  is 
\begin{eqnarray}
[Rim](t)=\frac{r_{im}}{k_c \gamma_{im}}\left[1-\frac{1}{A_S}-\frac{ \mid B_S\mid^{1/2}}{2 A_S^{3/2}} e^{-\gamma_{im} t} \log[( e^{\gamma_{im}t}-
\sqrt{\mid B_S\mid/A_S})/(e^{\gamma_{im}t}+\sqrt{\mid B_S\mid/A_S})]\right]+ c_{4S} e^{-\gamma_{im}t}.\label{eqricim2}
\end{eqnarray}
Assuming RicIp has a solution of the form 
$[Rip](t)=f_3(t) e^{-\gamma_{ip}t}$, we have 
the following differential equation satisfied by $f_3$
\begin{eqnarray}
\frac{df_3}{dt}= C_S e^{\gamma_{ip}t}-D_S e^{(\gamma_{ip}-\gamma_{im})t} {\Big(}\log[e^{\gamma_{im}t}-(\mid B_S\mid/A_S)^{1/2}]-\log[e^{\gamma_{im}t}+(\mid B_S\mid/A_S)^{1/2}]{\Big)}+E_S e^{(\gamma_{ip}-\gamma_{im})t},\label{eqf3}
\end{eqnarray}
where $C_S=r_{ip}R_0\frac{r_{im}}{k_c\gamma_{im}}(1-1/A_S)$, $D_S=r_{ip}R_0 \frac{r_{im} B_S^{1/2}}{ 2k_c \gamma_{im}A_S^{3/2}}$ and $E_S=r_{ip}R_0c_{4S}$.
Solving this equation, we finally find the solution for RicIp as  
\begin{eqnarray}
[Rip](t)=&& \frac{C_S}{\gamma_{ip}}+E_S \frac{e^{-\gamma_{im} t}}{\gamma_{ip}-\gamma_{im}}+\frac{D_S}{\gamma_{im}} 
e^{-\gamma_{ip}t}{\Big(} (e^{\gamma_{im}t}+(\mid B_S\mid/A_S)^{1/2})
\log[e^{\gamma_{im}t}+(\mid B_S\mid/A_S)^{1/2}]-\nonumber\\
&& (e^{\gamma_{im}t}-(\mid B_S\mid/A_S)^{1/2})
\log[e^{\gamma_{im}t}-(\mid B_S\mid/A_S)^{1/2}]{\Big)}+(c_{5S}-\frac{2D_S}{\gamma_{im}}(\mid B_S\mid/A_S)^{1/2})e^{-\gamma_{ip}t}, \label{eqricipa}
\end{eqnarray}where $c_{5S}$ is the integration constant. 

\begin{figure}
	\begin{minipage}[b]{0.35\textwidth}
		\includegraphics[width=\textwidth]{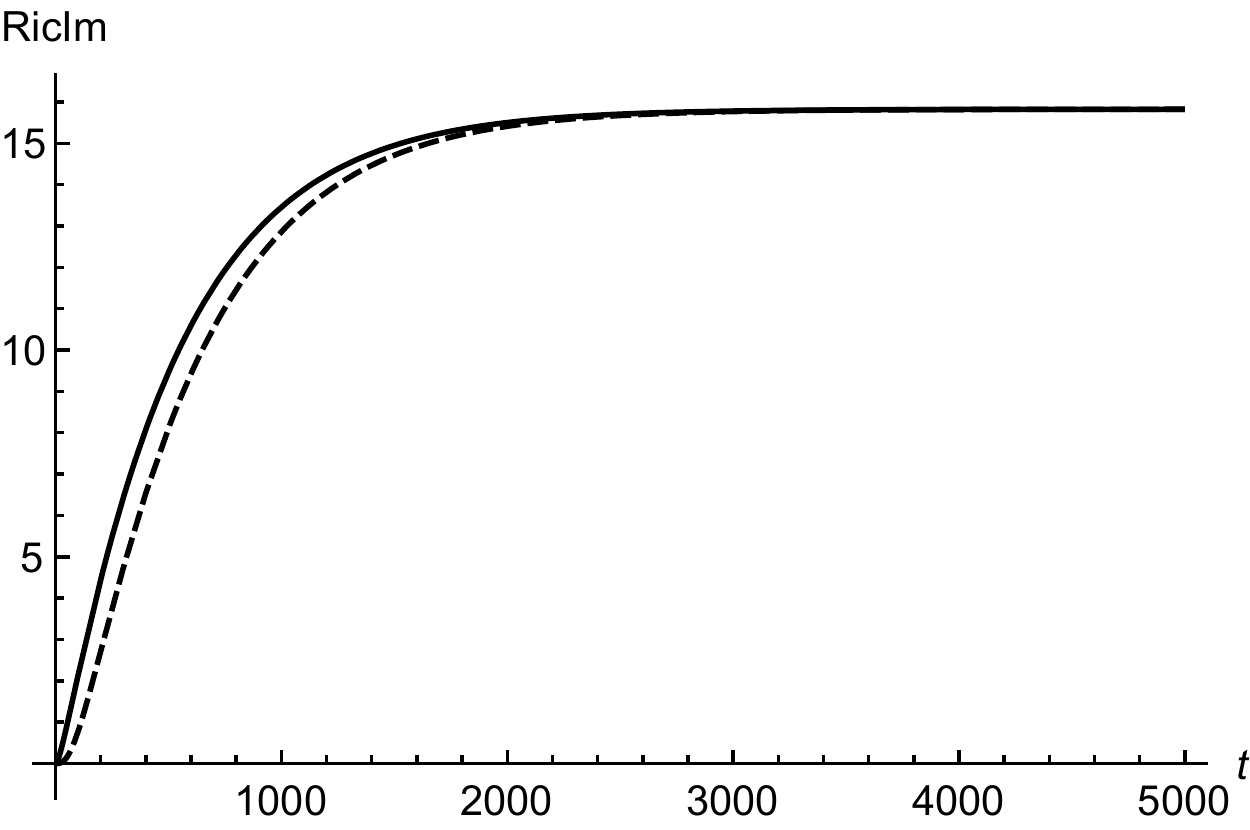} (A)
	\end{minipage}
	\begin{minipage}[b]{0.35\textwidth}
		\includegraphics[width=\textwidth]{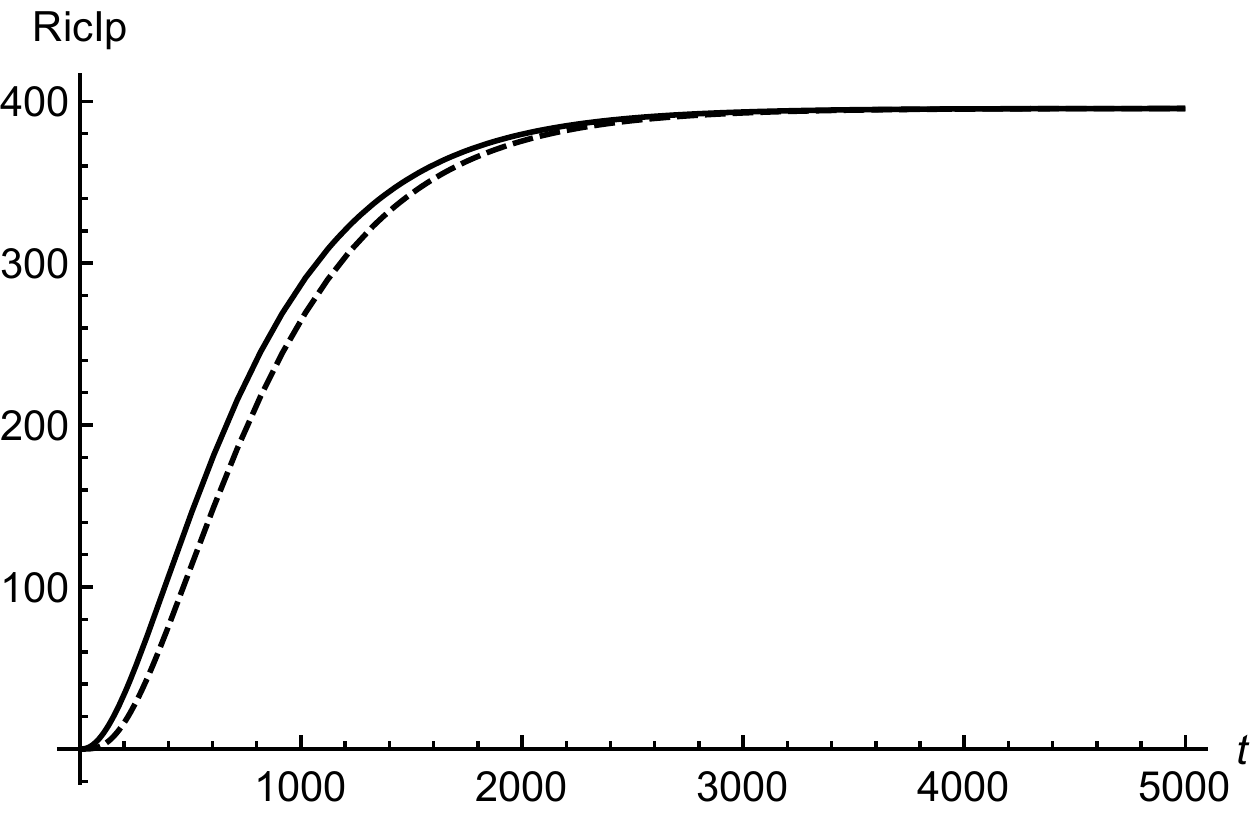} (B)
	\end{minipage}
	\caption{Dashed lines in both the figures represent numerical solutions. For these figures, all the synthesis rates such as $r_{\sigma m}$, $r_{\sigma p}$ etc. 
		have the same value, $0.01$ (${\rm  molecules.\ s^{-1}}$). 
		Further,  $\gamma_{im}=0.002$ (${\rm s}^{-1}$), $\gamma_{ip}=\gamma_{\sigma m}=0.004$ (${\rm s}^{-1}$),  $\gamma_{\sigma p}=0.004001$ 
		(${\rm s}^{-1}$)  ($\gamma_{\sigma m}\approx \gamma_{\sigma p}$) and $k_c=0.3$ (${\rm molecule}^{-1}$). 
		A constant concentration of $[Ra]=R_0=10$  molecules in the cell is assumed here. }
	\label{fig:math-numerics} 
\end{figure}

In order to obtain (\ref{eqricipa}), we have 
assumed $\gamma_{im}$ twice smaller than $\gamma_{ip}$. 
Considering initial concentrations of all the regulatory molecules to be negligible, 
we have  compared numerical and  mathematical solutions for RicIm and RicIp in 
figure (\ref{fig:math-numerics}). The numerical solutions provide an exact picture as how various 
concentrations vary with time.
Based on $\gamma_{\sigma p}\approx\gamma_{\sigma m}$ and $\gamma_{im}$ is 
approximately twice smaller than $\gamma_{\sigma p}$ or $\gamma_{\sigma m}$, we have obtained an approximate form of the 
integrand in equation (\ref{eqricim1}).   Due to this  approximation, the 
mathematical solutions deviate slightly from the exact, numerical solutions. 
Since the deviations are small, we assume  that (\ref{eqricim2}) and (\ref{eqricipa}) 
provide a reasonable quantitative description 
of time-variation in the target protein concentrations. 

\subsection{Under time-varying  concentration of sRNA}
Here, we find analytical solutions of  the  coupled, non-linear  equations  in (1) and   (2) of the main text. 
Unlike the previous section,
Rpra concentration is non-constant and it is governed by the equation
\begin{eqnarray}
&&\frac{d}{dt}[Ra]=r_a-\gamma_a [Ra], \ \ \ 
\end{eqnarray}
The solutions for $[Ra](t)$ and $[\sigma^sm](t)$ are as follows.
\begin{eqnarray}
[Ra](t)=c_1\  e^{-\gamma_a t}+\frac{r_a}{\gamma_a}, \ \ {\rm and}\ \ 
[\sigma^sm](t)=c_2\  e^{-\gamma_{\sigma m}t}+\frac{r_{\sigma m}}{\gamma_{\sigma m}}.
\end{eqnarray}
With these equations, the equation for $\sigma^s$ protein is as follows
\begin{eqnarray}
\frac{d}{dt}[\sigma^sp](t)=r_{\sigma p}\  (c_1 \ e^{-\gamma_a t}+\frac{r_a}{\gamma_a})
(c_2\  e^{-\gamma_{\sigma m}t}+\frac{r_{\sigma m}}{\gamma_{\sigma m}})-\gamma_{\sigma p} [\sigma^s p].
\end{eqnarray}
The solution for $[\sigma^s p](t)$  is
\begin{eqnarray}
[\sigma^s p](t) =c_3\  e^{-\gamma_{\sigma p}t}+A+B\  e^{-\gamma_a t}+C\ e^{-\gamma_{\sigma m}t} +  D\  e^{-\gamma_a t-\gamma_{\sigma m}t},
\end{eqnarray}
where various constants are as given below.
\begin{eqnarray}
&&A=\frac{r_{\sigma p} }{\gamma_{\sigma p}} \frac{r_a}{\gamma_a}\frac{r_{\sigma m}}{\gamma_{\sigma m}}, \ \ B=\frac{r_{\sigma p} }{(\gamma_{\sigma p}-\gamma_a)}\frac{r_{\sigma m}}{\gamma_{\sigma m}} ([Ra0]-\frac{r_a}{\gamma_a}), \ \ 
C=\frac{r_a}{\gamma_a}\frac{r_{\sigma p}}{(\gamma_{\sigma p}-\gamma_{\sigma m})} ([\sigma^s m0]-\frac{r_{\sigma m}}{\gamma_{\sigma m}}), \ \  {\rm and}\\ &&
D=\frac{r_{\sigma p}}{\gamma_{\sigma p}-(\gamma_a+\gamma_{\sigma m})} \ ([Ra0]-\frac{r_a}{\gamma_a}) \ ([\sigma^sm0]-\frac{r_{\sigma m}}{\gamma_{\sigma m}}),
\end{eqnarray} with $[\sigma^sm0]$ and $[Ra0]$ being the initial concentrations of Rpra and $\sigma^sm$.
The equation for RicIm can be solved by assuming a solution of the form 
$[Rim](t)=e^{-\gamma_{im} t} f_2(t)$ with $f_2(t)$ satisfying the equation 
\begin{eqnarray}
\frac{df_2(t)}{dt}=e^{\gamma_{im}t}\frac{r_{im}\sigma^sp}{1+k_c \sigma^sp}.
\end{eqnarray}
The solution for $f_2(t)$ is
\begin{eqnarray}
f_2(t)&=&\frac{r_{im}}{k_c}\int dt \ e^{\gamma_{im}t} (1-\frac{1}{1+k_c \sigma^sp}) 
\nonumber\\ 
&=&\frac{r_{im}}{k_c}\frac{e^{\gamma_{im}t}}{\gamma_{im}}-
\frac{r_{im}}{\gamma_{im}k_c} \int du\{1+
k_c[c_3\ u^{\frac{-\gamma_{\sigma p }}{\gamma_{im}}}+A+B\ u^{-\frac{\gamma_a}{\gamma_{im}}}+C\ u^{\frac{-\gamma_{\sigma m}}{\gamma_{im}}}+D u^{-(\frac{\gamma_a+\gamma_{\sigma m}}{\gamma_{im}})}]\}^{-1},
\end{eqnarray}
where $u=e^{\gamma_{im} t}$. 

\begin{figure}[h!]
	\begin{minipage}[b]{0.24\textwidth}
		\includegraphics[width= \textwidth]{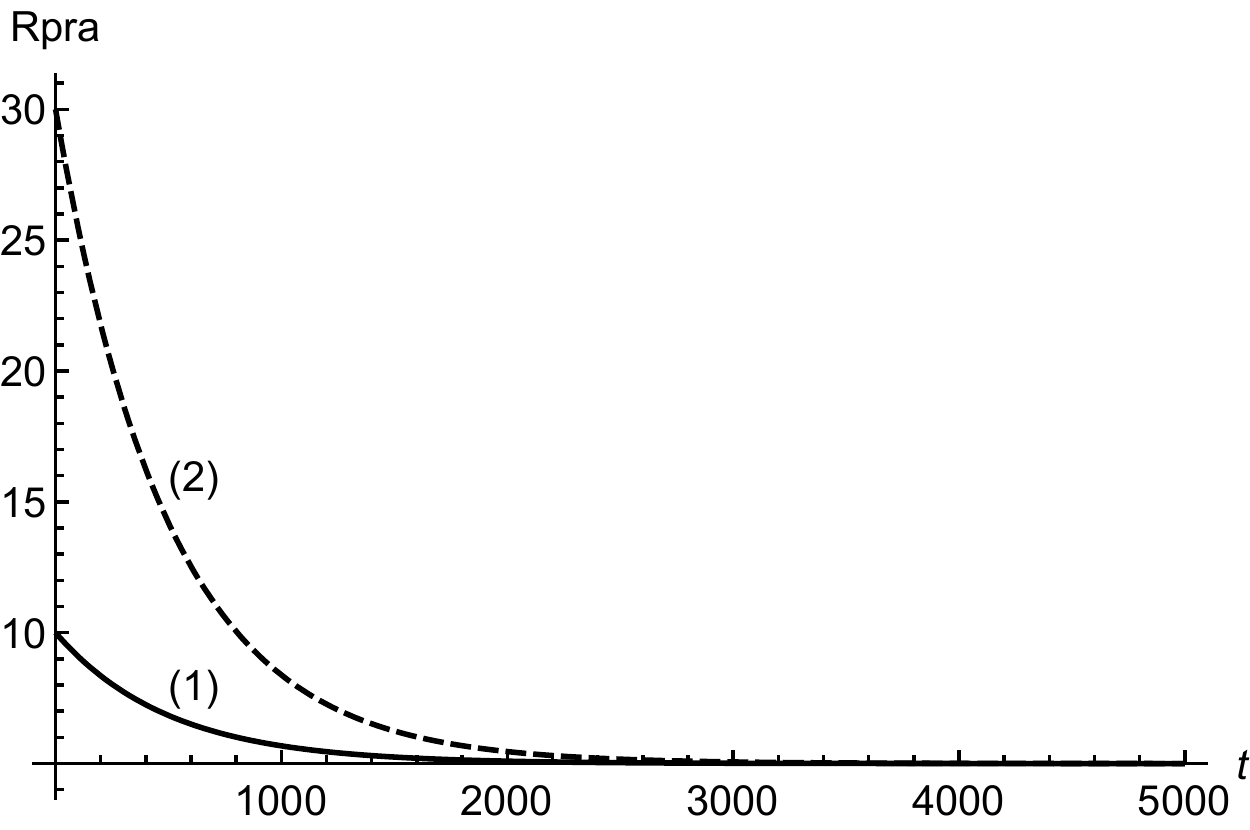} (A)
	\end{minipage}
	\begin{minipage}[b]{0.24\textwidth}
		\includegraphics[width=\textwidth]{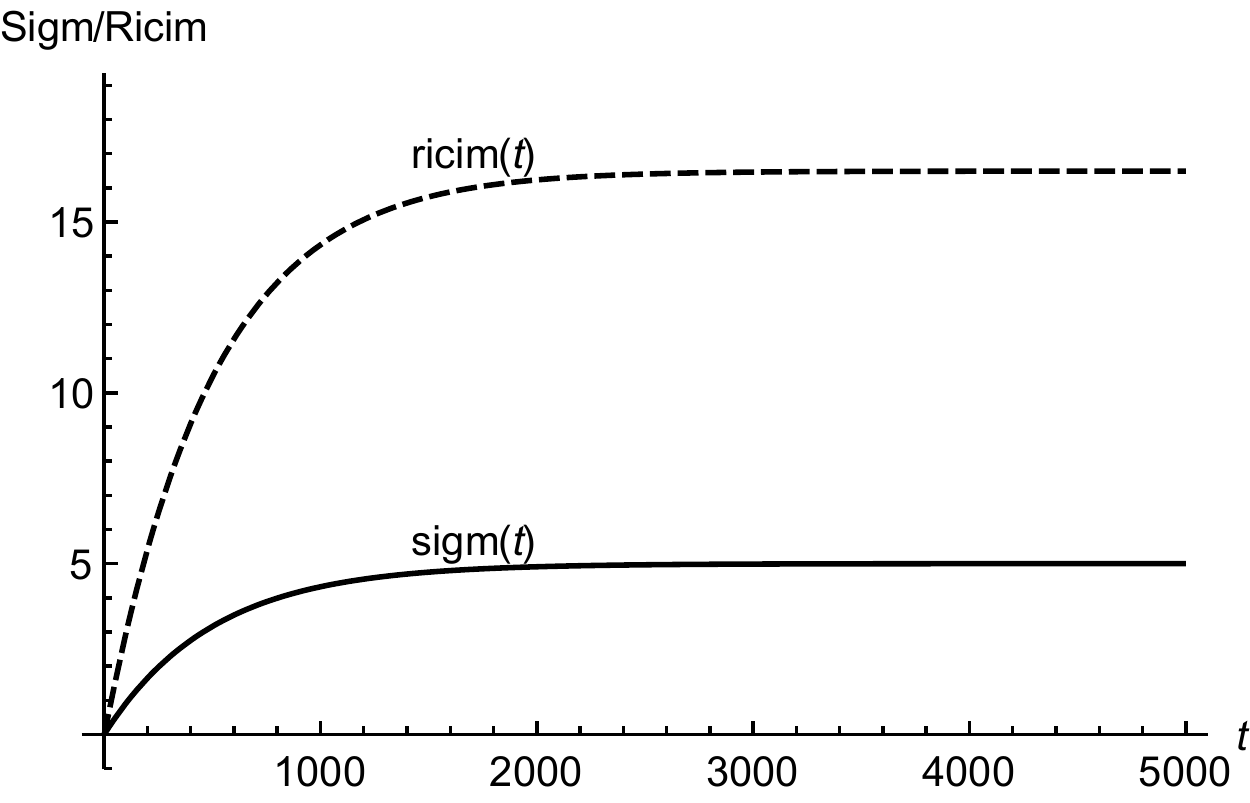} (B)
	\end{minipage}
	\begin{minipage}[b]{0.24\textwidth}
		\includegraphics[width=\textwidth]{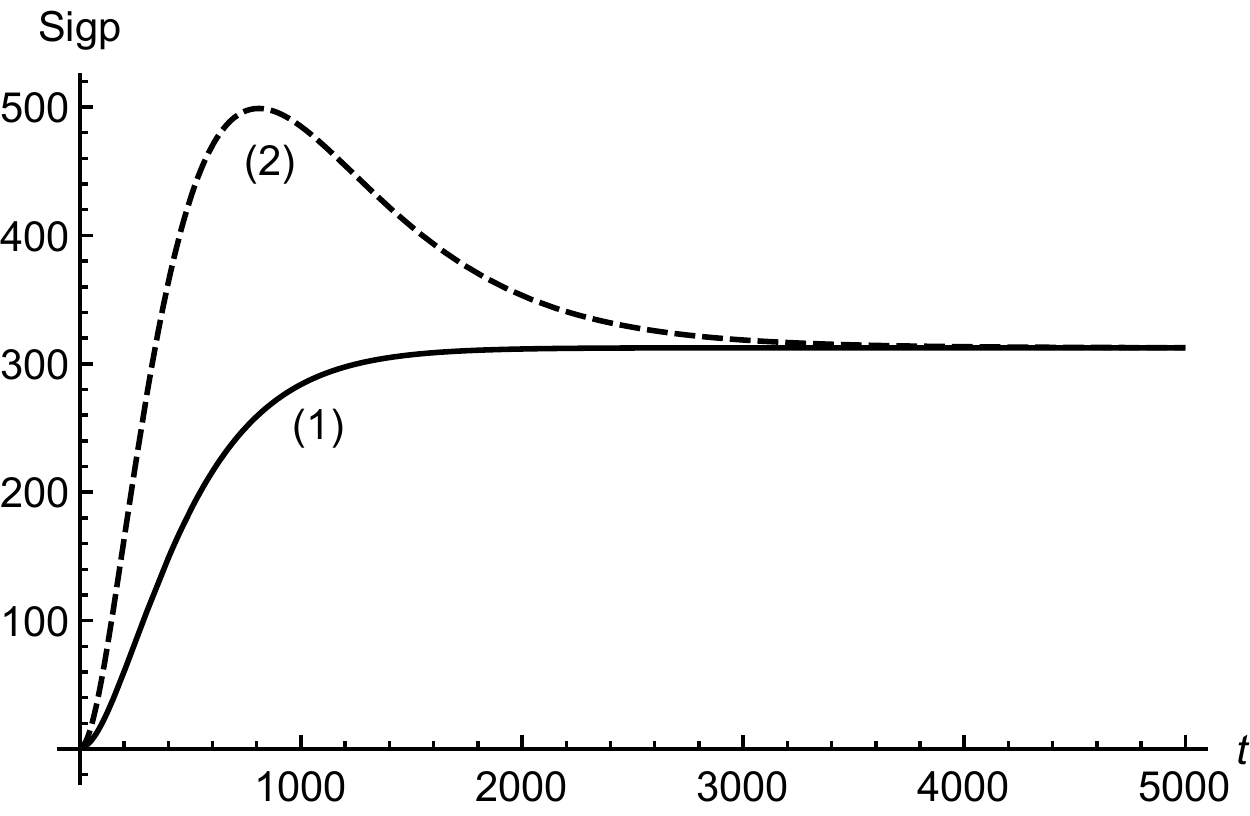} (C)
	\end{minipage}
	\begin{minipage}[b]{0.24\textwidth}
		\includegraphics[width=\textwidth]{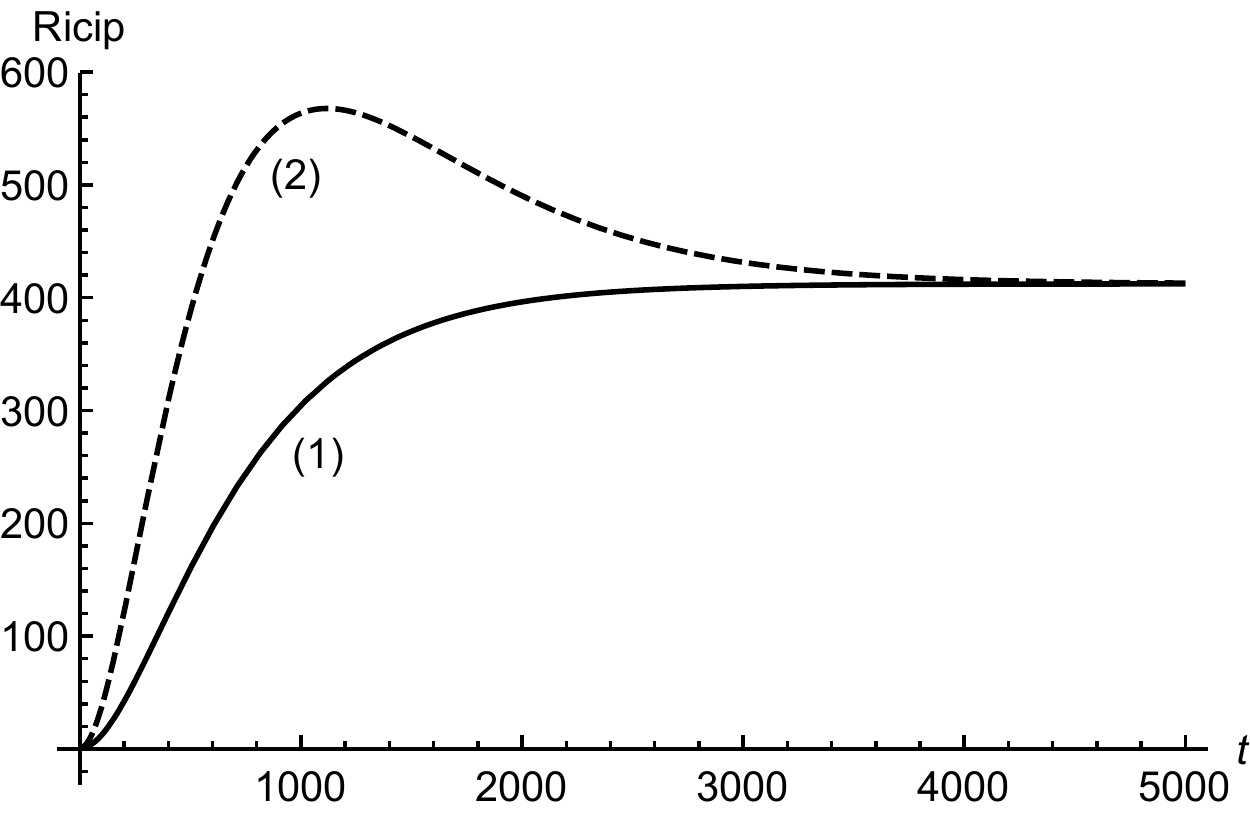} (D)
	\end{minipage}
	\caption{Mathematical solutions are plotted for two different initial conditions for RprA: (1) $[Ra]=10$ molecules, (2) $[Ra]=30$ molecules in the cell. 
		Initial numbers  for all the other regulators are chosen to zero. 
		All the synthesis rates are $0.01$ (${\rm  molecules.\ s^{-1}}$),   $\gamma_{im}=0.002$ (${\rm s}^{-1}$), $\gamma_a=\gamma_{ip}=\gamma_{\sigma m}=0.004$ (${\rm s}^{-1}$), 
		$\gamma_{\sigma p}=0.004001$ (${\rm s}^{-1}$) and and $k_c=0.3$ (${\rm molecule}^{-1}$). For  (B),  the plots for $\sigma^s_m$ and RicIm are almost the same for two different initial conditions.  These solutions are in complete agreement with direct  numerical solutions 
		of equations in (1) and (2) of the main text.  }
	\label{fig:generalfig}
\end{figure}

In order to have explicit  mathematical solutions for different concentrations, 
we assume  $\gamma_{im}=\gamma_a=\gamma_{\sigma m}$ and $\gamma_{im}\approx \frac{1}{2} \gamma_{\sigma p}$. Under these conditions, 
the integration for $f_2(t)$ can be done exactly.  We find 
\begin{eqnarray}
f_2(t)&=&\frac{r_{im}}{k_c\gamma_{im}} e^{\gamma_{im}t}  -\frac{r_{im}}{\gamma_{im} k_c} \int du \frac{u^2}{(Ak_c+1)u^2+(B+C)k_c u+k_c(D+c_3)}\nonumber\\
&=&\frac{r_{im}}{k_c\gamma_{im}} e^{\gamma_{im}t}  -\frac{r_{im}}{\gamma_{im} k_c}\Bigg[\frac{u}{(Ak_c+1)}-\frac{(B+C)k_c}{2(Ak_c+1)^2} \log[\mid(Ak_c+1)u^2+(B+C)k_c u+(D+c_3)k_c\mid]\nonumber\\
&+&\frac{1}{2(1+A k_c)^2}\frac{(-2k_c(D+c_3)(Ak_c+1)+(B+C)^2 k_c^2)}{\sqrt{ (B+C)^2 k_c^2-4(A k_c+1)(D+c_3)k_c}}\times\nonumber\\
&&\log{\bigg\{}\frac{2(Ak_c+1)u+(B+C)k_c-\sqrt{(B+C)^2 k_c^2-4(Ak_c+1)(D+c_3)k_c}}{2(Ak_c+1)u+(B+C)k_c+\sqrt{(B+C)^2 k_c^2-4(Ak_c+1)(D+c_3)k_c}}\bigg{\}} \Bigg]+c_4,
\end{eqnarray}
when $(B+C)^2 k_c^2-4(Ak_c+1)(D+c_3)>0$. 
The solution for RicI mRNA is thus 
\begin{eqnarray}
[Rim](t)&=&\frac{r_{im}}{k_c\gamma_{im}} -\frac{r_{im}}{\gamma_{im} k_c}\bigg[\frac{1}{(Ak_c+1)}-\frac{(B+C)k_c e^{-\gamma_{im} t}}{2(Ak_c+1)^2} 
\log[\mid(Ak_c+1)e^{2\gamma_{im}t}+(B+C)k_c e^{\gamma_{im}t}+(D+c_3)k_c\mid]\nonumber\\
&+&\frac{e^{-\gamma_{im}t}}{2(1+A k_c)^2}\frac{(-2k_c(D+c_3)(Ak_c+1)+(B+C)^2 k_c^2)}{\sqrt{(B+C)^2 k_c^2-4(A k_c+1)(D+c_3)k_c}}\times\nonumber\\
&&\log{\bigg\{}\frac{2(Ak_c+1)e^{\gamma_{im}t}+(B+C)k_c-\sqrt{(B+C)^2 k_c^2-4(Ak_c+1)(D+c_3)k_c}}{2(Ak_c+1)e^{\gamma_{im}t}+(B+C)k_c+\sqrt{(B+C)^2 k_c^2-4(Ak_c+1)(D+c_3)k_c}}\bigg{\}} \Bigg]+c_4e^{-\gamma_{im}t}\label{rimfinala}
\end{eqnarray}
for  $(B+C)^2 k_c^2-4(Ak_c+1)(D+c_3)>0$.
To find $[Rip](t)$, we numerically integrate the equation 
\begin{eqnarray}
\frac{d}{dt} [Rip](t)=r_{ip} (c_1\ e^{-\gamma_a t}+\frac{r_a}{\gamma_a})[Rim](t)-\gamma_{ip} [Rip](t)
\end{eqnarray}
using the mathematical solution for $[Rim](t)$ in (\ref{rimfinala}). 
The change in various concentrations are shown in figure (\ref{fig:generalfig}). 

\section{Moments}\label{sec:app3}
In the following, we show  our results for  first, second, and third moments derived using the generating function formulation described 
in sub-section (\ref{subsec:generate}) of the main text. 
\begin{eqnarray}
G_1 &=& \frac{r_s}{\gamma_s}\\
G_2 &=& \frac{r_{m_1}}{\gamma_{m_1}}\\
G_3 &=& \frac{r_{p_1} G_{12}}{\gamma_{p_1}}\\
G_4 &=& \frac{r_{m_2}^0 + r_{m_2}^1 G_3 }{\gamma_{m_2}}\\
G_5 &=& \frac{r_{p_2} G_{14}}{ \gamma_{p_2}}\\
G_{11} &=& \frac{r_s G_1 }{\gamma_s}\\
G_{22} &=& \frac{r_{m_1} G_2 }{\gamma_{m_1} }\\
G_{33} &=& \frac{r_{p_1} G_{123}}{ \gamma_{p_1}}\\
G_{44} &=& \frac{r_{m_2}^0 G_4 + r_{m_2}^1 G_{34}}{\gamma_{m_2}}\\
G_{55} &=&\frac{r_{p_2} G_{145}}{ \gamma_{p_2} }\\
G_{12}&=&\frac{r_s G_2 + r_{m_1} G_1 }{\gamma_s+\gamma_{m_1}}\\
G_{13}&=&\frac{r_s G_3 + r_{p_1}( G_{12}+G_{112} )}{\gamma_s + \gamma_{p_1}}\\
G_{14}&=&\frac{r_s G_4 + r_{m_2}^0 G_1+ r_{m_2}^1 G_{13}}{\gamma_s  +\gamma_{m_2}}\\
G_{15}&=&\frac{r_s G_5 + r_{p_2}( G_{14} + G_{114})}{\gamma_s +\gamma_{p_2}}\\
G_{23}&=&\frac{r_{m_1} G_3 + r_{p_1} (G_{12}+G_{122})}{\gamma_{m_1} +\gamma_{p_1}}\\
G_{24}&=&\frac{r_{m_1} G_4 + r_{m_2}^0 G_2 + r_{m_2}^1 G_{23}}{\gamma_{m_1} +\gamma_{m_2}}
\end{eqnarray}
\begin{eqnarray}
G_{25}&=&\frac{r_{m_1} G_5 + r_{p_2} G_{124}}{\gamma_{m_1} +\gamma_{p_2}}\\
G_{34}&=&\frac{r_{p_1} G_{124}+ r_{m_2}^0 G_3+ r_{m_2}^1 ( G_3 + G_{33})}{\gamma_{p_1} +\gamma_{m_2}}\\
G_{35}&=&\frac{r_{p_1} G_{125} + r_{p_2} G_{134}}{\gamma_{p_1} +\gamma_{p_2}}\\
G_{45}&=&\frac{r_{m_2}^0 G_5 + r_{m_2}^1 G_{35} + r_{p_2}( G_{14} + G_{144})}{\gamma_{m_2} + \gamma_{p_2}}\\
G_{111}&=& \frac{r_s G_{11}}{\gamma_s}\\
G_{112}&=& \frac{2 r_s G_{12} + r_{m_1} G_{11}}{2 \gamma_s +\gamma_{m_1}}\\
G_{113}&=& \frac{2 r_s G_{13} + 2 r_{p_1} G_{112} + r_{p_1} G_{1112}}{2 \gamma_s +\gamma_{p_1}}\\
G_{114}&=& \frac{2 r_s G_{14} +r_{m_2}^0 G_{11} +r_{m_2}^1 G_{113}}{ 2 \gamma_s + \gamma_{m_2}}\\
G_{115}&=& \frac{2 r_s G_{15} + r_{p_2} (2 G_{114} + G_{1114})}{2 \gamma_s + \gamma_{p_2}}\\
G_{122}&=& \frac{r_s G_{22} + 2  r_{m_1} G_{12}}{\gamma_s +2 \gamma_{m_1}}\\
G_{123}&=& \frac{r_s G_{23} + r_{m_1} G_{13} + r_{p_1} (G_{12} + G_{112} + G_{122} + G_{1122})}{\gamma_s +\gamma_{m_1} +\gamma_{p_1}}\\
G_{124}&=& \frac{r_s G_{24} + r_{m_1} G_{14} + r_{m_2}^0 G_{12} + r_{m_2}^1 G_{123}}{\gamma_s + \gamma_{m_1} + \gamma_{m_2}}\\
G_{125}&=& \frac{r_s G_{25} + r_{m_1} G_{15} + r_{p_2} (G_{124} +  G_{1124})}{\gamma_s + \gamma_{m_1} + \gamma_{p_2}}\\
G_{133}&=& \frac{r_s G_{33} + 2  r_{p_1}( G_{123}+ G_{1123} )}{\gamma_s +2 \gamma_{p_1}}\\
G_{134}&=& \frac{r_s G_{34} + r_{p_1} (G_{124} + G_{1124})+ r_{m_2}^0 G_{13} + r_{m_2}^1(G_{13}+ G_{133} ) }{\gamma_s + \gamma_{p_1} + \gamma_{m_2}}\\
G_{135}&=& \frac{r_s G_{35} + r_{p_1}( G_{125} +  G_{1125}) + r_{p_2} (G_{134} + G_{1134})}{\gamma_s + \gamma_{p_1} + \gamma_{p_2}}\\
G_{144}&=& \frac{r_s G_{44} + 2( r_{m_2}^0 G_{14} +  r_{m_2}^1 G_{134})}{\gamma_s +2 \gamma_{m_2} }\\
G_{145}&=& \frac{r_s G_{45} + r_{m_2}^0 G_{15} + r_{m_2}^1 G_{135} +r_{p_2} (G_{14}+ G_{114} +G_{144}+ G_{1144})}{\gamma_s  +\gamma_{m_2}+ \gamma_{p_2}}
\end{eqnarray}

\section{Translational activation by an sRNA }\label{sec:app4}
\begin{figure*}[ht]
	\centering
	\includegraphics[width=0.25\linewidth, height=0.15\textheight]{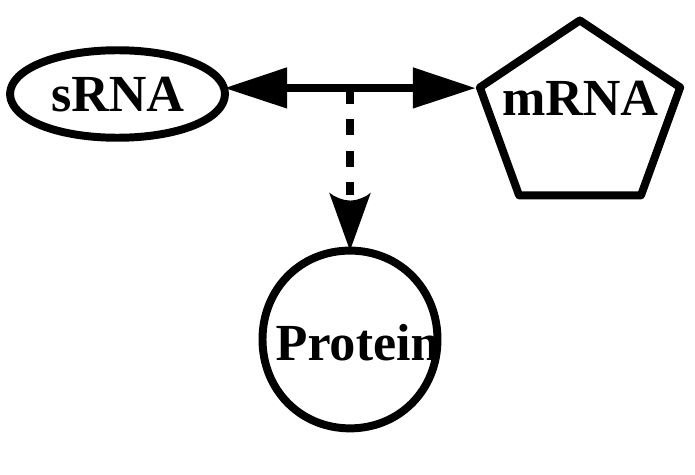}(A)
	\includegraphics[width=0.3\linewidth, height=0.15\textheight]{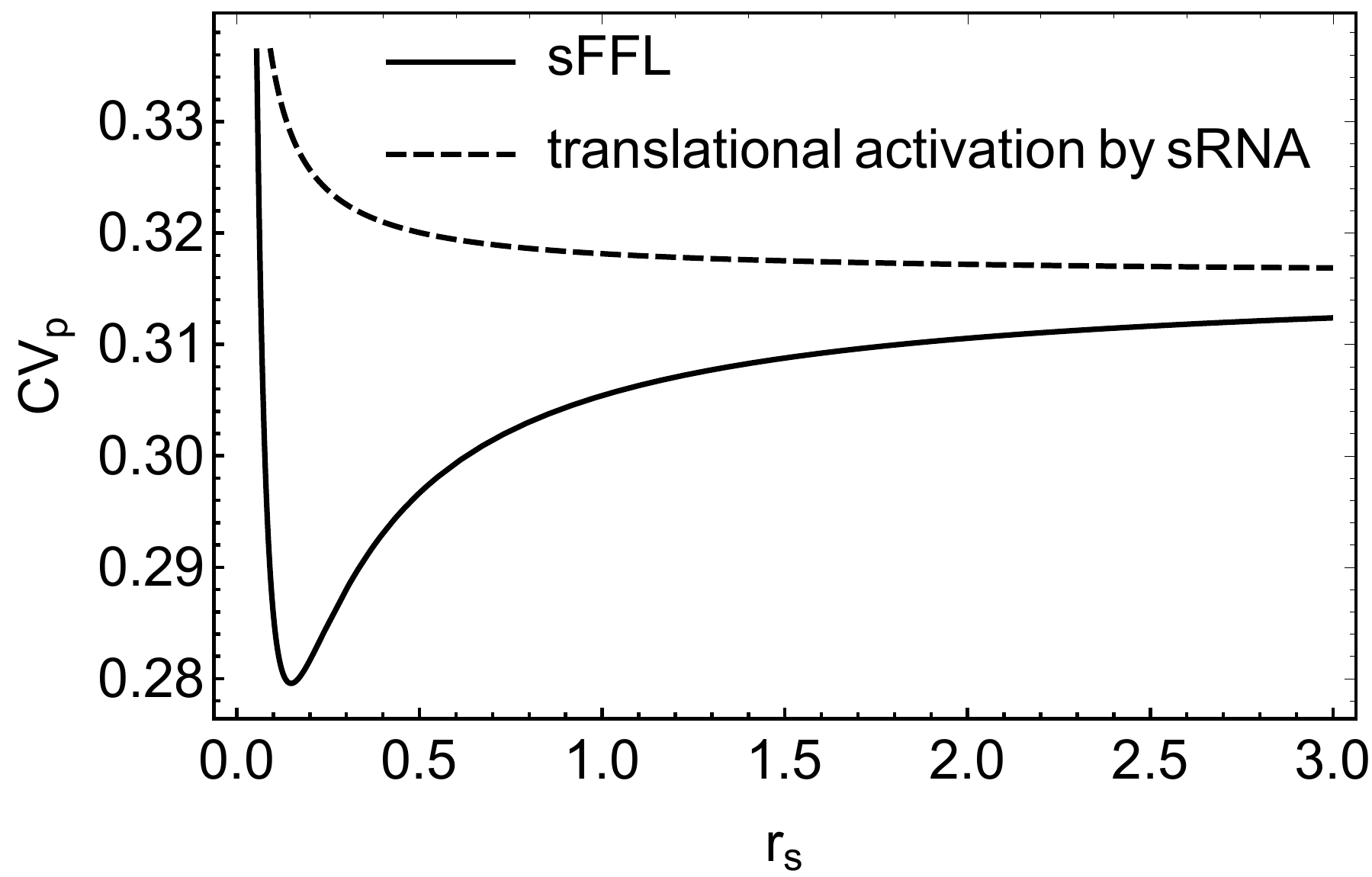}(B)
	\caption{ (A) Regulation of protein synthesis through  translational activation by sRNA. The dashed line with arrow indicates  translational activation by sRNA. (B) The coefficients of variation of the target  protein number for the regulation scheme of (A)  and for sFFL  are  plotted with the   synthesis rate ($r_s$) of sRNA.  For gene regulation with only translational activation, we have chosen synthesis and degradation rate of all components as $0.01\ (\rm molecules.\ s^{-1})$ and $0.002\  (\rm s^{-1})$, respectively. For sFFL, the parameter values are as mentioned in figure (\ref{fig:fluc})  with $\gamma_{p_1}=0.15\ (\rm s^{-1})$. }
	\label{fig:srna-regulation}
\end{figure*}

Here we consider a simple mechanism where the gene expression only involves translational activation by sRNA. In other words, sRNAs bind to the mRNAs and activate
translation (see Fig. (\ref{fig:srna-regulation}A)).  The  differential equations describing  variations of different concentrations with time are given below.
\begin{eqnarray}
&&\frac{d}{dt}[s]=r_s-\gamma_s [s],\\
&&\frac{d}{dt}{[m]}=r_m-\gamma_{m}[m],\ \text{and}\\
&&\frac{d}{dt}[p]=r_p [s][m]-\gamma_{ p} [p],
\end{eqnarray}
where $[s]$, $[m]$, and $[p]$ denote concentrations of sRNA, mRNA, and protein, respectively.
The stochastic analysis can be performed in a similar way as described in the main text. In this case, the number of moments required to find the coefficient of 
variation is smaller  compared to that 
 of sFFL and one requires the moments $G_{1}$, $G_{2}$, $G_{3}$, $G_{11}$, $G_{12}$, $G_{13}$, $G_{22}$, $G_{23}$, $G_{33}$, $G_{112}$, $G_{122}$, 
 $G_{123}$ and $G_{1122}$ to find the coefficient of variation. The coefficient of variation for the target protein  has been 
 plotted in figure (\ref{fig:srna-regulation}B).  

\section{Reaction Scheme}\label{sec:app5}
Here, we list different biochemical reactions considered in  stochastic simulations.  The values of various rate constants used   for figure (\ref{fig:fluca}A) 
are mentioned inside the brackets. 

\begin{eqnarray}
&&\schemestart $\phi$ \arrow{->[$r_s$]} $s$ \schemestop\quad ({\rm supply\ of\  sRNA;}\  r_s={\rm variable}) 
\\
&&\schemestart $s$ \arrow{->[$\gamma_{s}$]} $\phi$ \schemestop\quad ({\rm degradation\ of\ sRNA;}\  \gamma_s=0.002\ (\rm s^{-1})) 
\\
&&\schemestart $\phi$ \arrow{->[$r_{ m_1}$]} $m_1$ \schemestop\quad ({\rm supply\ of\  mRNA\ m_1;}\  r_{m_1}= 0.01\  (\rm molecules.\ s^{-1})) 
\\
&&\schemestart $m_1$ \arrow{->[$\gamma_{ m_1}$]} $\phi$\schemestop\quad  ({\rm degradation\ of\  mRNA,\ m_1;}\  \gamma_{m_1}=0.002\ (\rm s^{-1}))
\\
&&\schemestart $m_1$ + $s$ \arrow{->[$k_a^+$]} $m_1\mbox{-}s$ \schemestop\quad  ({\rm mRNA - sRNA\ complex\ formation;}\  k_a^+=0.01\  (\rm molecules^{-1}.\ s^{-1})) 
\\
&&\schemestart $m_1$-$s$ \arrow{->[$k_a^-$]} $m_1$ + $s$ \schemestop\quad  ({\rm mRNA - sRNA\ complex\ dissociation;}\  k_a^-=0.01\  (\rm s^{-1})) 
\\
&&\schemestart $m_1$-$s$ \arrow{->[$r_{p_1}$]} $p_1$ + $m_1$-$s$ \schemestop\quad ({\rm translation\ and\  synthesis \ of\   protein\ p_1;\  r_{p_1}=0.01\  (\rm molecules.\ s^{-1})}\ )  
\\
&&\schemestart $p_1$ \arrow{->[$\gamma_{p_1}$]}  $\phi$\ \schemestop\quad ({\rm degradation\ of\ protein\  p_1;}\ \gamma_{p_1}=0.05\ (\rm s^{-1}),\ 0.15\ (\rm s^{-1}))   
\\
&&\schemestart $p_1$ + $G_I$ \arrow{->[$k_c^+$]}   $G_I^*$\ \schemestop\quad  ({\rm transcriptional\ activation\ of\  gene\ synthesising\ m_2;}\  k_c^+=0.2\  (\rm molecules^{-1}.\ s^{-1}))
\\
&&\schemestart $G_I^*$  \arrow{->[$k_c^-$]} $p_1$ + $ G_I$ \schemestop\quad ({\rm deactivation\ of\ gene\ synthesising\ m_2;}\  k_c^-=2\ (\rm s^{-1})) 
\\
&&\schemestart $G_I^*$ \arrow{->[$r_{m_2}$]} $m_2$ +  $G_I^*$\schemestop\quad ({\rm synthesis\ of\  mRNA\ m_2;}\  r_{m_2}= 0.005\  (\rm molecules.\ s^{-1}))  
\\
&&\schemestart $m_2$ \arrow{->[$\gamma_{m_2}$]} $\phi$\schemestop\quad  ({\rm degradation\ of\ mRNA, m_2;}\ \gamma_{m_2}=0.002\ (\rm s^{-1})) 
\\
&&\schemestart $m_2$ + $s$ \arrow{->[$k_i^+$]} $m_2$-$s$  \schemestop\quad ({\rm mRNA - sRNA\ complex\ formation;}\ k_i^+=0.005\  (\rm molecules^{-1}.\ s^{-1}))
\\
&&\schemestart $m_2$-$s$ \arrow{->[$k_i^-$]} $m_2 + s$  \schemestop\quad ({\rm mRNA - sRNA\ complex\ dissociation;}\ k_i^-=0.005\ (\rm s^{-1})) 
\\
&&\schemestart $m_2$-$s$ \arrow{->[$r_{p_2}$]} $p_2$ + $m_2$-$s$ \schemestop\quad  ({\rm translation\ and\ synthesis \ of\   protein\  p_2;}\ r_{p_2}=0.01\  (\rm molecules.\ s^{-1}))
\\
&&\schemestart $p_2$ \arrow{->[$\gamma_{p_2}$]}  $\phi$\schemestop\quad ({\rm degradation\ of\ protein\ p_2}; \ \gamma_{p_2}=0.002\ (\rm s^{-1}))
\end{eqnarray}

Here, $G_I$ and $G_I^*$ denote the inactivated and activated  form of the  gene synthesising $m_2$.  For figure (\ref{fig:fluca}B), we choose $r_s=0.01\ (\rm molecules.\ s^{-1})$.

\end{document}